\documentclass[12pt]{article}
\usepackage[english]{babel}    
\usepackage[ansinew]{inputenc} 
\usepackage[T1]{fontenc}       
\usepackage{lmodern,microtype} 
\usepackage{amsmath,amsthm,amssymb,amsfonts} 
\usepackage{dsfont,mathrsfs,ushort} 
\usepackage{titlesec,titling}  
\usepackage[nohead]{geometry}  
\usepackage{setspace,caption}  
\usepackage{enumitem,booktabs} 
\usepackage[comma,longnamesfirst]{natbib} 
\usepackage{pstricks,pst-all}   
\usepackage{tikz}
\usepackage{pst-plot,pst-node,pst-3dplot}
\usepackage{hyperref,breakurl}
\usepackage{pgfplots}
\usepackage{graphicx}
\usepackage{caption}
\usepackage{subcaption}
\usepackage{longtable}
\usepackage[flushleft]{threeparttable}
\usepackage{algorithmic}
\setenumerate{label=\small(\roman*)}
\hypersetup{colorlinks=true,pdfnewwindow=true,pdfstartview=FitH,%
	pdftitle="Peer Selection",pdfauthor="Nail Kashaev and Natalia Lazzati",%
	urlcolor=blue!90!red!45!black,citecolor=blue!90!red!45!black,linkcolor=red!90!black}
\DeclareCaptionFont{fancy}{\bfseries\sffamily}
\allowdisplaybreaks

\providecommand{\psreset}{\psset{%
		linewidth=0.3pt,linestyle=solid,linecolor=black,
		dotsize=2.5pt,dotsep=2.5pt,arrowsize=4pt,
		fillstyle=none,fillcolor=white,
		showpoints=false,arrows=-,linearc=0,framearc=0,
		hatchsep=2pt,hatchwidth=0.2pt,nodesep=4pt,opacity=1}
	\psset{gridcolor=black!60, subgridcolor=black!30}
}

\psreset

\usepackage{graphicx}
\graphicspath{ {Figures/overall/}{Figures/last_period_top50/} }
\titleformat{\section}[block]{\centering\large\bfseries\sffamily}{\thesection.}{0.5em}{}
\titleformat{\subsection}[block]{\flushleft\bfseries}{\thesubsection.}{0.5em}{}
\titleformat{\subsection}[block]{\flushleft\bfseries\sffamily}{\thesubsection.}{0.5em}{}
\titleformat{\subsubsection}[runin]{\normalsize\itshape}{\bfseries\upshape\sffamily\thesubsubsection.}{0.5em}{}[.--\:]
\renewcommand{\thesubsubsection}{\arabic{section}.\arabic{subsection}.\alph{subsubsection}}
\titlespacing{\section}{0ex}{10ex}{5ex}
\titlespacing{\subsection}{0in}{6ex}{3ex}
\titlespacing{\subsubsection}{0mm}{2ex}{0.5em}
\providecommand{\abstitle}[1]{{\par\vspace*{2ex}\small\bfseries\sffamily #1}\hspace*{1ex}}
\renewenvironment{abstract}%
{\begin{center}\begin{minipage}{0.8\linewidth}%
			\setlength{\parindent}{0.0em}\abstitle{Abstract}\small}%
		{\end{minipage}\end{center}\vfill\clearpage}

\newtheorem{proposition}{Proposition}[section]

\DeclareMathOperator*{\argmax}{arg\,max}

\providecommand{\tr}{^{\prime}}

\newcommand{\Char}[1]{\mathds{1}\left(\,#1\,\right)}

\newcommand{\abs}[1]{\left\lvert#1\right\rvert}

  \theoremstyle{remark}
  \newtheorem{rem}{\protect\remarkname}
  \theoremstyle{plain}
  
  \theoremstyle{definition}
  
\theoremstyle{plain}

  \theoremstyle{plain}
  
 \theoremstyle{definition}
  \newtheorem{example}{\protect\examplename}
  \theoremstyle{plain}
  \newtheorem{assumption}{\protect\assumptionname}

  \providecommand{\assumptionname}{Assumption}

  \providecommand{\definitionname}{Definition}
  \providecommand{\lemmaname}{Lemma}
  
  \providecommand{\remarkname}{Remark}
\providecommand{\corollaryname}{Corollary}
\providecommand{\theoremname}{Theorem}
\providecommand{\examplename}{Example}

\usepackage{xcolor}

\newcounter{aux}
\newcounter{eg1}

\title{Discrete Choice with Endogenous Peer Selection\thanks{For their useful comments we thank Victor Aguiar, Roy Allen, Tim Conley, and David Rivers. We would also like to thank the audience at Network Science and Economics conference (Miami Herber BS) and Triangle Econometrics Conference (UNC), and the seminar presentations at Duke University. Kashaev gratefully acknowledges financial support from Social Sciences and Humanities Research Council Development and  Insight Development Grants.}
}

\author{ 
	Nail Kashaev\\
	Western University\\
	nkashaev@uwo.ca
	\and
	Natalia Lazzati\\
	UC Santa Cruz\\
	nlazzati@ucsc.edu
	}
\date\today

\begin{document}

\maketitle

\begin{abstract}
\noindent We develop a continuous time discrete choice model of peer effects. The distinctive feature of the model is that agents might not consider all peers at the moment of making a decision. Instead, they select \emph{some} of them on the basis of a mechanism that depends on recent choices. We characterize the equilibrium behavior and study the empirical content of the limited attention peer effect model. We allow changes in the choices of peers to affect both the set of peers to which the agent pays attention and her preferences over the alternatives. We exploit variation in choices together with variation in the size of the set of potential peers (or reference groups) to recover the preferences of the agents and the peer selection mechanisms. We apply our results to model expansion and contraction decisions by fast-food restaurants and find evidence of limited attention to actions of competitors.
\bigskip \bigskip


\end{abstract}
\section{Introduction}

\noindent Social interactions are a cornerstone of many economic and social outcomes \citep{durlauf2001social}. Consumers purchase products after seeing their friends buy them, and firms open new stores in response to the actions of their competitors. A large empirical literature measures these peer effects in discrete choice settings under the assumption that agents observe or pay attention to all their peers.\footnote{This paper uses a behavioral definition of peers. We think the peers of an agent are all other agents that can directly affect her choices.} This means that, for instance, consumers take into account the purchasing decisions of all their friends and firms consider the actions of all their competitors at the moment of making a decision. However, in many settings, different constraints or limitations might lead agents to pay attention to the choices of a smaller subset of peers, e.g., agents could be choosing actions using a subjective model of peer effects that may be misspecified. This paper offers a model of peer selection in which the peer selection mechanism is allowed to depend on recent choices.

Similarly to \citet{kashaev2019peer} and \citet{kashaev2025peer}, we study a continuous-time discrete choice model with social interactions. The network specifies the set of potential peers (referred to as reference group) of each agent. These connections are known by the agents, but not by the researcher. At random times governed by independent Poisson alarm clocks, an agent wakes up and selects an alternative from a finite menu. When the clock rings, the agent first draws a subset of peers, the \emph{active peer set} or \emph{active peers}, from her reference group. The probability of selecting any given peer depends on the current choices of the agent and her peers, allowing agents to pay more attention to peers that have selected similar alternatives. The agent evaluates the alternatives under the influence of the active peers and then selects an option. The resulting profile of choices evolves according to a continuous-time Markov process. We show that (under some restrictions) if the researcher observes the choices of all agents in the population over a long period of time, then she can recover the social network (i.e., the reference group for each agent), the peer selection mechanism, and the random preferences of each agent captured by the distribution over the choice set conditional on realization of the choices of the active peers.  

The modeling restrictions ensure the existence of a unique equilibrium. This equilibrium is characterized by an invariant distribution over the set of choice configurations, which describes the long-run frequencies of each possible choice profile. For the identification of the model, the important feature is that the invariant distribution has full support. Since the probability of selecting a peer is affected by previous choices of connected agents, which are determined by the equilibrium distribution, the realized distribution over random active peer sets is endogenous.

We illustrate key features of the model through a simple example with a closed-form solution. The example shows that allowing agents to include peers who choose similar alternatives more often in the active peer set increases the correlation of choices among agents with similar preferences. In specific applications, this result can be interpreted as polarization and our model can be seen as offering a micro-foundation for echo chambers \citep{ anqi2025}. 

Our contribution is threefold: First, we connect bounded rationality with peer effects by embedding endogenous peer selection into a discrete choice model of peer effects in preferences. In a simple two-agent example, we show how endogenous attention can change the comparative statics: paying attention to peers that make similar choices can either reinforce or dampen coordination relative to a standard model depending on whether peer effects are positive or negative. Second, we offer a simple setting with transparent identification arguments that can be incorporated (as we explain below) into a larger model of either correctly specified or misspecified network structure. Similarly to the use of menu variation in consideration set models, we exploit variation in the sizes of the reference groups to identify the network, the peer-selection mechanism, and preference parameters from a single long panel of choices while imposing only mild exclusion and dimension-reducing conditions. Third, we show how the limited attention to firms operating in the same market can shape expansion and contraction decisions in our empirical illustration. 

The starting point of the identification strategy is the set of Conditional Choice Probabilities (CCPs). Each CCP specifies the frequency of choices made by an agent given the choice configuration at the time of making the decision. We assume the analyst can consistently estimate these CCPs from the data. We first show that, under a set of mild restrictions, variation in the choices of potential peers generates variation in the frequency of choices of a given agent. Thus, the set of potential peers, or reference groups, can be recovered from the data. We then implement a thought experiment: We show that if the researcher could manipulate the reference groups of the agents by creating or deleting links, then she could recover the rest of the model via a recursive strategy. However, in the applications that motivate this paper, the researcher has no control over reference groups and cannot arbitrarily alter them. We therefore replicate the same logic using two models in which the necessary variation arises endogenously within the model. The first model assumes there is variation in the sizes of reference groups across similar agents. If agents were people, this is the same as saying that some people have more friends than others. The second model, instead, assumes no peer effects in the outside option. This modeling restriction has been motivated by many empirical applications, where the outside option means \emph{do nothing}.

We recover the peer-selection mechanism and the choice rule of each agent nonparametrically as a function of the choice configurations which, in our dynamic setting, correspond to the states of the system. These two parts of the model are left quite general and can be easily microfounded with various existing results in the literature. For instance, we could generate the peer-selection mechanism as the product of random meetings and the probability of including the met agents in the active peer set based on relative payoffs ---that would capture the advantage of doing so. We could also view the choice rule as the reduced-form value from selecting an alternative. This reduced-form could capture forward-looking behavior without explicitly modeling it. In this case, the choice rule would implicitly reflect the immediate payoff of selecting the option and the discounted value of payoffs that occur in the continuation state. In this context, agents might display correct beliefs about the social environment or a rather misspecified model. 

We illustrate our model by studying the expansion and contraction decisions of five major fast-food restaurant chains operating in geographically distinct markets in China. In this setting, each firm-market pair is an agent and firms respond to the actions of their competitors when deciding whether to open or close restaurants. We allow managers to ignore the presence of some competitors. This can happen for different reasons. For example, the manager may not think that a particular restaurant is a close substitute at a specific moment in time. In addition, monitoring competitors is costly in terms of time, data collection, and analysis, and managers may be unable to process all relevant market information continuously. This idea builds on previous theories of bounded rationality in firms offered by \citet{simon1955behavioral}. Consistent with this view, we find substantial heterogeneity in attention to competitors across firms and markets. Limited attention leads firms to overexpansion, and for some firms it primarily induces distortions in the probabilities of opening stores in different cities, while for others it fundamentally changes the way they respond to the actions of competitors.

We finally connect our work to the existing literature. First, the ideas are related to the consideration set models where agents do not pay attention to all available alternatives \citep[see, for instance,][ and references therein]{aguiar2023random}. While this literature focuses on \emph{consideration sets of alternatives}, we study \emph{consideration sets of peers} in an interactive framework. In doing so, our work also contributes to the growing econometric literature on network formation \citep[see][for an overview]{graham2015methods, de2017econometrics, chandrasekhar2016econometrics}. Within this literature, the studies by \citet{leung2015econometric, menzel2015strategic, miyauchi2016structural, boucher2017my, mele2017structural, depaula2018identifying, thirkettle2019identification, ridder2020two, sheng2020networks, badev2021nash} and \citet{gualdani2021econometric} have analyzed game-theoretic models of creating links. Another related literature models social decision-making as based on observations from a sampled subset of network neighbors (e.g. \citealp{degroot74, bala98, golub10}) or from a small set of peers with its size constrained by the cost of processing the information \citep{park2026}. We complement these approaches by offering a general peer-selection mechanism in which the selection probabilities for effective network formation depend on recent choices. This distinction matters for identification, as endogenous active peer set formation can generate correlated behaviors. 

There is a large econometric literature on the identification of models of social interactions where the choices of peers shape agents' preferences over alternatives (see, for example, \citealp{blume2011identification, bramoulle2020peer, de2017econometrics} and \citealp{graham2015methods}, for comprehensive reviews of this literature). On the theoretical side, \citet{chambers2025} recover the impacts that agents have on each other using variation in group composition. Although existing social interaction models typically assume correct beliefs about peer behavior, a growing literature documents biased social learning and misperceived peer behavior, e.g., formation of peer-induced beliefs \citep{boucher22} and misclassification in social interactions \citep{lin24, meyer17, lewel25}. We can think of our model as a discrete choice framework with endogenous peer selection that incorporates misspecified beliefs about peers.\footnote{See \citet{esponda2016} for an equilibrium set-up of misspecified models.}

Finally, our empirical application relates to the literature on misoptimizing (e.g., boundedly rational) firm behavior in the behavioral industrial organization  \citep{simon1945administrative,armstrong2010behavioral, heidhues2018behavioral}. Most of this literature introduces deviations from rationality to the demand side of the economy (i.e. behavioral consumers). In contrast, we study a setting in which firms themselves may misoptimize \citep{bloom2007, dellavigna2019uniform, gold2011,  hortacsu2008,cho2010flat, tourek2022targeting, almunia2024strategic, dube2025monopsony, galdon2025value, song2025firm, hjort2026across, kashaev2025peer}.

The remainder of the paper is organized as follows. Section~\ref{sec: model} presents the model and its assumptions. Section~\ref{sec: equilibrum} elaborates on equilibrium behavior. We establish identification of the model parameters in Section~\ref{sec: identification}. Section~\ref{sec: application} presents our empirical application and we conclude in Section~\ref{sec: conclusion}. All proofs can be found in Appendix~\ref{app: proofs}.

\section{Model}\label{sec: model}

\noindent This section describes the model and the main assumptions used in the paper. 

\subsection{Network, Peer Selection, and Preferences}

\paragraph{Network} There is a finite set of agents $\mathcal{A}=\{1,2,\dots,A\}$, $A\geq2$, choosing from a finite set of alternatives $\mathcal{Y}=\left\{0, 1,\dots,Y\right\}$, $Y\geq1$. In some contexts, $0$ will indicate the default alternative. We refer to $\mathbf{y}=\left( y_{a}\right)_{a\in \mathcal{A}}\in \mathcal{Y}^{A}$ as a choice configuration. Agents are connected through a social network. The network is described by a set of edges. Each edge identifies two connected agents and the direction of the connection. For each Agent $a\in \mathcal{A}$, her reference group is defined as follows:
\begin{equation*}
\mathcal{N}_{a}=\left\{ a'\in \mathcal{A}:a'\neq a\text{
and there is an edge from }a\text{ to }a'\right\}. 
\end{equation*}

\setcounter{eg1}{\value{example}}
\begin{example}
Suppose that there are four agents and three alternatives. That is, $\mathcal{A}=\left\{ 1,2,3,4\right\}$ and $\mathcal{Y}=\left\{0,1,2\right\}$. The reference groups are
\[
\mathcal{N}_{1}=\left\{2, 3\right\},\quad \mathcal{N}_{2}=\left\{1\right\},\quad \mathcal{N}_{3}=\left\{2\right\}, \quad\mathcal{N}_{4}=\emptyset.
\]
This means that, for instance, Agents 2 and 3 may affect the choices of Agent 1.
\hfill $\square$
\end{example}

\paragraph{Peer Selection and Preferences} The revision of choices follows a standard continuous-time Markov process. Agents have independent Poisson alarm clocks with rates $\mathbf{\lambda }=\left( \lambda_{a}\right)_{a\in \mathcal{A}}$. The alarm of Agent $a$ is triggered at exponentially distributed moments with mean $1/\lambda_{a}$. When this happens, the agent first selects a subset of peers $\mathcal{N}\subseteq\mathcal{N}_a$ to whom she will pay attention (i.e., the active peers) and then makes a choice under their influence (i.e., only the active peers affect her current choice). We do not model the cognitive process that leads to peer selection. We take it as given and identify it nonparametrically. Many existing models can be used to motivate the peer-selection mechanism we invoke (see Section~\ref{sec: ps} for details). (All identification results in Section~\ref{sec: modelid} are valid if agents make simultaneous decisions, i.e., they have the same clock.)

For each agent in the population, the described setting introduces a two-stage decision process that depends on the current configuration of choices ($\mathbf{y}$):

\noindent \textbf{Step 1} Agent $a$ picks a set of active peers $\mathcal{N}\subseteq\mathcal{N}_a$. The ex-ante probability that set $\mathcal{N}$ is picked is
$
\operatorname{S}^{a}\left(\mathcal{N} \mid \mathbf{y}, \mathcal{N}_a \right)
$
with $\operatorname{S}^{a}\left(\mathcal{N} \mid \mathbf{y}, \mathcal{N}_a \right) \geq 0$ and $\sum\nolimits_{\mathcal{N}\subseteq\mathcal{N}_a}\operatorname{S}^{a}\left(\mathcal{N} \mid \mathbf{y}, \mathcal{N}_a \right) = 1$.

\noindent $\textbf{Step 2}$ After selecting peers $\mathcal{N}\subseteq\mathcal{N}_a$, the agent picks an alternative according to a choice rule
$
\operatorname{R}^{a}\left( \cdot \mid \mathbf{y}, \mathcal{N}\right)
$
that satisfies $\operatorname{R}^{a}\left( v \mid \mathbf{y}, \mathcal{N}\right) > 0$ for all $v$ and $\sum\nolimits_{{v}\in\mathcal{Y}}\operatorname{R}^{a}\left(v \mid \mathbf{y}, \mathcal{N}\right) = 1$. 

The choice rule summarizes the decision process after the peer selection is completed. We associate choice rules with preferences, since, in many instances, the preference parameters can be recovered from the choice rules. If agents are forward-looking, the choice rule will also contain information about beliefs on future actions of other agents and discounted values of future payoffs (see Section~\ref{sec: rc} for further details).

\setcounter{aux}{\value{example}}
\setcounter{example}{\value{eg1}}
\begin{example}[continued]
Suppose that, at the moment of making a choice, Agent $1$ only selects Agent $2$ from her reference group $\mathcal{N}_a=\{2,3\}$. That is $\mathcal{N}=\{2\}$. Hence, her probability of selecting alternative 1 is
$
    \operatorname{R}^{1}\left(1  \mid \mathbf{y},\left\{2\right\}\right).
$ 
Assume Agent $a$'s indirect utility from $v$ given the active peer set $\mathcal{N}$ is $u_{a,v}\left(\mathbf{y},\mathcal{N}\right)+\xi_{a,v,\mathcal{N}}$, where $u_{a,v}$ captures the mean utility from the alternative. The vector of agent- and peer-group-specific taste shocks $(\xi_{a,v,\mathcal{N}})_{v\in\mathcal{Y}}$ is continuously distributed with conditional cumulative distribution function (c.d.f.) $\operatorname{F}_{a,\xi}(\cdot\mid \mathbf{y},\mathcal{N})$. Then, for $v\in\mathcal{Y}$
\[
\operatorname{R}^1\left(v  \mid \mathbf{y},\left\{2\right\}\right)=\int\Char{v=\argmax_{v'\in\mathcal{Y}}\{u_{a,v'}\left(\mathbf{y},\left\{2\right\}\right)+\xi_{a,v',\{2\}}\}}d\operatorname{F}_{a,\xi}((\xi_{a,v,\{2\}})_{v\in\mathcal{Y}}\mid  \mathbf{y},\left\{2\right\}).
\]
If the $\xi_{a,v,\mathcal{N}}$s are independent and identically distributed (i.i.d.) shocks, distributed according to the standard Type I extreme value distribution, then the above expression simplifies to
\[ \quad \quad \quad \quad \quad \quad \quad \quad \quad  \quad \quad
    \operatorname{R}^{1}\left(v  \mid \mathbf{y},\left\{2\right\}\right)=\dfrac{\exp\left(u_{1,v}\left(\mathbf{y},\left\{2\right\}\right)\right)}{\sum_{v'\in\mathcal{Y}}\exp\left(u_{1,v'}\left(\mathbf{y},\left\{2\right\}\right)\right)}.\quad \quad \quad  \quad \quad \quad \quad  \quad \quad  \quad \quad  \quad  \hfill \square \]
\end{example}
As the example demonstrates, our framework allows dependence between preferences and selection of peers. That is, the peer effect in preferences can appear both via mean utility (e.g., direct effect from playing the same video game with your friends) and latent preference shocks (e.g., learning the latent quality of the product via choices of peers).

At the moment of making a choice, the ex-ante probability for Agent $a$ of selecting each alternative is a finite mixture given by 
\begin{equation*}
\operatorname{P}_{a}\left( v \mid \mathbf{y}\right) =\sum\nolimits_{\mathcal{N} \subseteq \mathcal{N}_{a}}\operatorname{R}^{a}\left(v \mid \mathbf{y}, \mathcal{N}\right)\operatorname{S}^{a}\left(\mathcal{N} \mid \mathbf{y}, \mathcal{N}_a \right).
\end{equation*}
\noindent We refer to $\operatorname{P} = \left(\operatorname{P}_a\right)_{a\in\mathcal{A}}$ as the set of Conditional Choice Probabilities (CCPs).

\subsection{Main Assumptions}

\noindent This section adds more structure to the peer selection process and preferences. Regarding the selection process, we assume that each peer is selected independently from the others. 
\begin{assumption}[Independent Selection]\label{ass: MM}
    The probability of considering a subset of peers $\mathcal{N} \subseteq \mathcal{N}_a$ is given by
    \[
    \operatorname{S}^a\left(\mathcal{N} \mid \mathbf{y}, \mathcal{N}_a \right) = \prod_{a'\in\mathcal{N}}\operatorname{Q}^a(a'\mid \mathbf{y})\prod_{a'\in\mathcal{N}_a\setminus\mathcal{N}}(1-\operatorname{Q}^a(a'\mid \mathbf{y})),
    \]
    where $\operatorname{Q}^a(a'\mid \mathbf{y})$ is the probability that Agent $a'$ is selected by Agent $a$ given $\mathbf{y}$ such that $\operatorname{Q}^a(a'\mid \mathbf{y})=0$ for $a'\not\in\mathcal{N}_a$ and $0<\operatorname{Q}^a(a'\mid \mathbf{y})\leq 1$ for $a'\in\mathcal{N}_a$. 
\end{assumption}
\noindent Assumption~\ref{ass: MM} reduces the dimensionality of the identification problem. Without it, for each $\mathbf{y}$, the researcher would need to learn $\operatorname{S}^a$ over $2^{\abs{\mathcal{N}_a}}$ points. Assumption~\ref{ass: MM} reduces this number to $\abs{\mathcal{N}_a}$. This assumption may hold in some settings, but not in others. It is violated if, for example, adding Agent $a'$ to the active peer set affects the likelihood that Agent $a''$ also be added. Assumption~\ref{ass: MM} also requires that every peer in each reference group be selected with positive probability. 

We also impose restrictions on $\operatorname{R}^a$ to make the problem more tractable. Let $\operatorname{N}_a^v(\mathbf{y},\mathcal{N})$ be the number of peers of Agent $a$ in the nonempty set $\mathcal{N} \subseteq \mathcal{N}_a$ who pick $v$ in $\mathbf{y}$. That is,
\[
\operatorname{N}_a^v(\mathbf{y},\mathcal{N})=\sum_{a'\in\mathcal{N}}\Char{y_{a'}=v}.
\]
We let $\operatorname{N}_a(\mathbf{y},\mathcal{N})=(\operatorname{N}_a^v(\mathbf{y},\mathcal{N}))_{v\in\mathcal{Y}}$ be the vector of numbers of active peers selecting each alternative. We let $\operatorname{N}_a(\mathbf{y},\emptyset)=\mathbf{0}$ since, for any non-empty $\mathcal{N}$, $\operatorname{N}_a(\mathbf{y},\mathcal{N})$ has at least one non-zero component.

\begin{assumption}\label{ass: simple}
    For each $a$, $\mathbf{y}$, $a'\in\mathcal{N}_a$, $\mathcal{N}\subseteq\mathcal{N}_a$, $\mathcal{N}\neq\emptyset$, and $v$
    \begin{enumerate}
        \item $\operatorname{Q}^a(a'\mid \mathbf{y})= \operatorname{Q}_{a}(a'\mid y_a,y_{a'})$; and
        \item $\operatorname{R}^a(v\mid\mathbf{y},\mathcal{N})=\operatorname{R}_a(v\mid y_a,\operatorname{N}_a(\mathbf{y},\mathcal{N}))$.
    \end{enumerate}
\end{assumption}
\noindent Assumption~\ref{ass: simple}(i) states that the selection probability of a specific peer depends on the current choice of the agent and the one of the target peer. It also depends on the identity of the target peer. This specification allows agents to pay more (or less) attention to peers who have chosen similar alternatives in the recent past. The choices of other peers do not affect this selection probability.   

\setcounter{aux}{\value{example}}
\setcounter{example}{\value{eg1}}
\begin{example}[continued]
Assume that Agent $a$ follows a threshold rule when selecting peers to pay attention to: Agent $a'$ is selected (at the moment of making a choice) if and only if $c_{a,a'}\left(y_a,y_{a'}\right) \geq \varepsilon$, where $c_{a,a'}\left(y_a,y_{a'}\right)$ is an agent-pair-specific attention index that depends on the previous choices of both agents and $\varepsilon$ is a random attention shock independent of $\mathbf{y}$. Then, the probability of selecting Agent $a'$ is
    $
    \operatorname{Q}_{a}\left(a'\mid y_{a}, y_{a'} \right) = 
    \operatorname{F}_{\varepsilon}\left(c_{a,a'}\left(y_a,y_{a'}\right)\right),
    $
    where $\operatorname{F}_{\varepsilon}$ is the c.d.f. of $\varepsilon$, and Assumption~\ref{ass: simple}(i) is satisfied.
    \hfill $\square$
\end{example}

Assumption~\ref{ass: simple}(ii) requires the choice rules to depend on the number of active peers selecting each alternative and not on the identity of peers. However, we allow for unrestricted dependence of choice rules on agent's own choice $y_a$. 

\setcounter{aux}{\value{example}}
\setcounter{example}{\value{eg1}}
\begin{example}[continued]
Suppose that indirect utilities of alternatives are linear-in-means or linear-in-aggregates: $u_{a,v}(\mathbf{y},\mathcal{N})=\alpha_{a,v}(y_a)+\beta_{a,v}(y_a)\operatorname{N}^v_{a}(\mathbf{y},\mathcal{N})/\abs{\mathcal{N}}$ or $u_{a,v}(\mathbf{y},\mathcal{N})=\alpha_{a,v}(y_a)+\beta_{a,v}(y_a)\operatorname{N}^v_{a}(\mathbf{y},\mathcal{N})$, where $\alpha_{a,v}(y_a)$ and $\beta_{a,v}(y_a)\neq 0$ are unknown parameters that depend on the current choice of Agent $a$. Then Assumption~\ref{ass: simple}(ii) is satisfied.
\hfill $\square$
\end{example}

The last assumption in this section is a mild regularity condition that rules out certain knife edge cases. Specifically, it ensures that when a peer switches her choice, the resulting effect on peer selection does not exactly offset the effect on preferences. This assumption allows us to identify the set of potential peer groups for each agent. Let $\mathbf{0}^v_{a'}$ be the configuration obtained from the zero vector by replacing the $a'$-th component by $v$. That is, this is a configuration where every one but Agent $a'$ picked the default and Agent $a'$ picked alternative $v$.

\begin{assumption}[Regularity]\label{ass: regularity} For all $a\in\mathcal{A}$, $a'\in\mathcal{N}_a$, there exists $v$ such that
        \[
        \dfrac{\operatorname{Q}_{a}(a'\mid 0,v)}{\operatorname{Q}_{a}(a'\mid 0,0)}\neq \dfrac{\sum_{\mathcal{N}\subseteq\mathcal{N}_a\setminus\{a'\}}\left[\operatorname{R}^{a}\left(v \mid \mathbf{0}, \mathcal{N}\cup\{a'\}\right)-\operatorname{R}^{a}\left(v \mid \mathbf{0}, \mathcal{N}\right)\right]\operatorname{S}^{a}\left(\mathcal{N} \mid \mathbf{0}, \mathcal{N}_a\setminus\{a'\} \right)}{\sum_{\mathcal{N}\subseteq\mathcal{N}_a\setminus\{a'\}}\left[\operatorname{R}^{a}\left(v \mid \mathbf{0}^v_{a'}, \mathcal{N}\cup\{a'\}\right)-\operatorname{R}^{a}\left(v \mid \mathbf{0}, \mathcal{N}\right)\right]\operatorname{S}^{a}\left(\mathcal{N} \mid \mathbf{0}, \mathcal{N}_a\setminus\{a'\} \right)}.
        \]
\end{assumption}
The left hand side of Assumption~\ref{ass: regularity} captures the change in the probability that a peer is selected when she switches from the default option $0$ to option $v$. The right hand side captures the corresponding change in the preference component of the choice rule induced by the peer's action. Assumption~\ref{ass: regularity} requires that these two effects are not equal. The restriction is imposed only at one choice configuration where everyone picks the default $0$. This is not necessary, we only need the existence of a point where peer effects in preferences and in selection do not offset each other. Moreover, this choice configuration can vary with $a$ and $a'$. For example, if switching Agent $a'$ to an alternative increases or decreases both $\operatorname{R}^a$ and $\operatorname{Q}^a$, then Assumption~\ref{ass: regularity} is automatically satisfied.

\subsection{Microfoundations for the Peer Selection Mechanism and Random Choices}
\noindent In the paper, we take the process of peer selection and the choice rules as given and recover them nonparametrically. Next, we show that these processes can be generated by some well-known empirical models. 

\subsubsection{Peer Selection Mechanism}\label{sec: ps}
Let us say that Agent $a$ creates a link (at the moment of choosing) to Agent $a' \in \mathcal{N}_a$ if she includes this other agent in her active peer set. The selection of active peers is then governed by the probabilities of links between Agent $a$ and all other agents in $\mathcal{N}_a$. There are many standard models for link probabilities based on random utility, limited attention, or continuous-time matching frictions (see, e.g., \citealp{duffie2005,jackson2008}, and \citealp{jackson1996}). In these models, the links often result from random meetings and payoffs: When two agents randomly meet, forming a link yields a deterministic surplus that depends on agents' states and past behavior. In addition, agents face idiosyncratic costs or frictions that are privately observed. The link forms when the surplus is larger than the cost. We could interpret our model as summarizing this two-outcome process using a reduced-form selection probability. We allow the components of the link probability to depend on the previous actions of Agent $a$ and the other agents in her reference group.  

To illustrate this idea, let
\begin{equation*}
\delta_{a,a'}\left(y_a,y_{a'}\right)  \in \left(0,1\right)
\end{equation*}
be the probability that Agent $a$ meets Agent $a' \in \mathcal{N}_a$ at the moment of decision making. When Agents $a$ and $a'$ meet, the acceptance decision is governed by a random utility representation. Specifically, if Agent $a$ meets Agent $a'$, then the former obtains utility 
\begin{equation*}
\operatorname{U}_{a,a'}\left(y_a,y_{a'}\right) = u_{a,a'}\left(y_a,y_{a'}\right) + \varepsilon
\end{equation*}
\noindent where $u_{a,a'}$ is a deterministic surplus function and $\varepsilon$ is an idiosyncratic Logistic random shock. Agent $a$ decides to add Agent $a'$ to the set of active peers if $\operatorname{U}_{a,a'}\left(y_a,y_{a'}\right) \geq 0$. This implies the selection probability
\begin{equation*}
\operatorname{Q}_{a}\left(a'\mid y_a,y_{a'}\right) = \delta_{a,a'}\left(y_a,y_{a'}\right) \frac{\exp\left(u_{a,a'}\left(y_a,y_{a'}\right)\right)}{1 + \exp\left(u_{a,a'}\left(y_a,y_{a'}\right)\right)}.
\end{equation*}
In this setup, the probability of paying attention to peers in $\mathcal{N}\subseteq \mathcal{N}_a$ is given by
 \[
    \operatorname{S}^a\left(\mathcal{N} \mid \mathbf{y}, \mathcal{N}_a \right) = \prod_{a'\in\mathcal{N}}\frac{\delta_{a,a'}\left(y_a,y_{a'}\right) \exp\left(u_{a,a'}\left(y_a,y_{a'}\right)\right)}{1 + \exp\left(u_{a,a'}\left(y_a,y_{a'}\right)\right)}\prod_{a'\in\mathcal{N}_a\setminus\mathcal{N}}\frac{1+\left(1-\delta_{a,a'}\left(y_a,y_{a'}\right) \right)\exp\left(u_{a,a'}\left(y_a,y_{a'}\right)\right)}{1 + \exp\left(u_{a,a'}\left(y_a,y_{a'}\right)\right)}.
    \]
Randomness in the peer selection mechanism might also come from agents having a wrong model of who they should pay attention to or how the network works. Among other possibilities, they could misperceive who is relevant, use a wrong rule for sampling peers, or misspecify the selection probabilities. In these setups, peer selection reflects errors in beliefs, not just noise.

\subsubsection{Choice Rule}\label{sec: rc} 
The random choice rule in our setting could simply reflect best choices for preference rankings with peer effects. Alternatively, while we do not model forward-looking behavior explicitly, one could view the choice rule as the reduced-form value from selecting an alternative. This reduced-form would implicitly incorporate the immediate payoff of selecting the option and the discounted value of payoffs that occur in the continuation state. An explicit forward-looking setup would also require adding beliefs (about future choices) to the model. Although this alternative formulation can lead to multiple equilibria, the empirical literature on dynamic games often assumes that there is a unique equilibrium in the data-generating process \citep{blevins2018identification}. 

In our model, it is also possible that agents would be choosing actions using a subjective model of peer effects that may be misspecified. In this case, the random components of the utility represent the prediction errors generated by the misspecification.

\section{Equilibrium Behavior}\label{sec: equilibrum}

\noindent The independent Poisson processes of the alarm clocks of the agents lead the selection process of alternatives through time. They guarantee that the transition rate from choice configuration $\mathbf{y}$ to any different one $\mathbf{y}'$ is as follows
\begin{equation}
\operatorname{w}\left( \mathbf{y}' \mid \mathbf{y}\right) =\left\{ 
\begin{array}{lcc}
0 & \text{if} & \sum\nolimits_{a\in \mathcal{A}}\mathds{1}\left( y_{a}'\neq
y_{a}\right) >1 \\ 
\sum\nolimits_{a\in \mathcal{A}}\lambda_{a}\operatorname{P}_{a}\left( y_{a}' \mid \mathbf{y}\right) \mathds{1}\left( y_{a}'\neq y_{a}\right) & 
\text{if} & \sum\nolimits_{a\in \mathcal{A}}\mathds{1}\left( y_{a}'\neq
y_{a}\right) =1
\end{array}
\right. .  \label{T}
\end{equation}
These transition rates are the off-diagonal terms of the \textit{transition rate matrix}.\footnote{This transition rate matrix has many zeros in known locations. \citet{blevins2017identifying,blevins2018identification} offers a nice discussion of this feature and its advantage for identification over discrete time models.} The diagonal terms are constructed from them in a simple way
\begin{equation*}
\operatorname{w}\left( \mathbf{y}\mid \mathbf{y}\right) =-\sum\nolimits_{\mathbf{y}%
'\in \mathcal{Y}^{A}\setminus \left\{ \mathbf{y}\right\}
}\operatorname{w}\left( \mathbf{y}'\mid \mathbf{y}\right).
\end{equation*}

We will indicate by $\mathcal{W}$ the transition rate matrix. As the number of choice configurations is $\left(Y+1\right)^{A}$, it follows that $\mathcal{W}$ is a $\left(Y+1\right)^{A}\times \left(Y+1\right)^{A}$ matrix.  
An equilibrium in this model is an invariant distribution  $\mu: \mathcal{Y}^{A} \rightarrow\left[0,1\right]$, with $\sum\nolimits_{\mathbf{y}\in \mathcal{Y}^{A}}\mu \left( \mathbf{y}\right) =1$, of the dynamic process with transition rate matrix $\mathcal{W}$. It indicates the likelihood of each choice configuration $\mathbf{y}$ in the long run. This equilibrium behavior relates to the transition rate matrix in a linear fashion 
\begin{equation*}
\mu \mathcal{W}=\mathbf{0}.
\end{equation*}

The assumptions in the paper guarantee the existence and uniqueness of an invariant distribution. Importantly, this invariant distribution has full support (i.e., any choice configuration $\mathbf{y}$ realizes with positive probability).  

\begin{proposition}\label{prop: existence} If Assumptions~\ref{ass: MM} and~\ref{ass: simple} hold, there exists a unique, full support equilibrium $\mu$.
\end{proposition}

\begin{example} Suppose that the network consists of two identical agents that select between two alternatives, option $1$ and the default option $0$. Let us also assume that the random preferences only depend on the choice of the other agent. As agents are identical, we will drop $a$ from some terms. Thus, for $a=1,2$, the CCPs can be written in a rather succinct form 
\begin{equation*}
\operatorname{P}_{a}\left( v \mid y_1,y_2 \right) =\operatorname{Q}\left( y_a,y_{-a}\right) \operatorname{R}\left( v \mid 1-y_{-a},y_{-a}\right) + \left(1 - \operatorname{Q}\left(y_a,y_{-a}\right)\right)\operatorname{R}\left(v \mid 0,0\right).
\end{equation*}
\noindent The rates for their Poisson alarm clocks are $1$. The transition rate matrix $\mathcal{W}$ is as follows.
 
\begin{equation*}
\begin{tabular}{c|c|c|c|c|}
\multicolumn{1}{c}{}&\multicolumn{1}{c}{$\left(0,0\right)$}&\multicolumn{1}{c}{$\left(0,1\right)$}&\multicolumn{1}{c}{$\left(1,0\right)$}&\multicolumn{1}{c}{$\left(1,1\right)$}\\
\cline{2-5}
$\left( 0,0\right)$ & $*$ & $\operatorname{P}_{2}\left( 1 \mid 0,0\right)$ & $\operatorname{P}_{1}\left( 1 \mid 0,0\right)$ & $0$ \\ \cline{2-5}
$\left( 0,1\right)$ & $\operatorname{P}_{2}\left( 0 \mid 0,1\right)$ & $*$ & $0$ & $\operatorname{P}_{1}\left( 1 \mid 0,1\right)$ \\ \cline{2-5}
$\left( 1,0\right)$ & $\operatorname{P}_{1}\left( 0 \mid 1,0\right)$ & $0$ & $*$ & $\operatorname{P}_{2}\left( 1 \mid 1,0\right)$ \\ \cline{2-5}
$\left( 1,1\right)$ & $0$ & $\operatorname{P}_{1}\left( 0 \mid 1,1\right)$ & $\operatorname{P}_{2}\left( 0 \mid 1,1\right)$ & $*$ \\ \cline{2-5}
\end{tabular}%
\
\end{equation*}

\noindent The diagonal terms, *, are such that the elements in each line add up to $0$. Note that, by symmetry, we have $\operatorname{P}_{1}\left( 1 \mid 1,1\right) = \operatorname{P}_{2}\left( 1 \mid 1,1\right)$, $\operatorname{P}_{1}\left( 0 \mid 0,1\right) = \operatorname{P}_{2}\left( 0 \mid 1,0\right)$, $\operatorname{P}_{1}\left( 0 \mid 1,0\right) = \operatorname{P}_{2}\left( 0 \mid 0,1\right)$, and $\operatorname{P}_{1}\left( 0 \mid 0,0\right) = \operatorname{P}_{2}\left( 0 \mid 0,0\right)$. After a simple calculation, the invariant distribution of choices, or steady-state equilibrium, is given by:
\begin{align*}
    \mu_{(0,0)}&=\dfrac{\operatorname{P}_{1}\left( 0 \mid 1,0\right)\operatorname{P}_{1}\left( 0 \mid 1,1\right)}{\operatorname{P}_{1}\left( 1 \mid 0,0\right)\operatorname{P}_{1}\left( 1 \mid 0,1\right)+\operatorname{P}_{1}\left( 0 \mid 1,0\right)\operatorname{P}_{1}\left( 0 \mid 1,1\right)+2\operatorname{P}_{1}\left( 1 \mid 0,0\right)\operatorname{P}_{1}\left( 0 \mid 1,1\right)},\\
    \mu_{(1,0)}&=\mu_{(0,1)}=\dfrac{\operatorname{P}_{1}\left( 1 \mid 0,0\right)\operatorname{P}_{1}\left( 0 \mid 1,1\right)}{\operatorname{P}_{1}\left( 1 \mid 0,0\right)\operatorname{P}_{1}\left( 1 \mid 0,1\right)+\operatorname{P}_{1}\left( 0 \mid 1,0\right)\operatorname{P}_{1}\left( 0 \mid 1,1\right)+2\operatorname{P}_{1}\left( 1 \mid 0,0\right)\operatorname{P}_{1}\left( 0 \mid 1,1\right)},\\
    \mu_{(1,1)}&=\dfrac{\operatorname{P}_{1}\left( 1 \mid 0,0\right)\operatorname{P}_{1}\left( 1 \mid 0,1\right)}{\operatorname{P}_{1}\left( 1 \mid 0,0\right)\operatorname{P}_{1}\left( 1 \mid 0,1\right)+\operatorname{P}_{1}\left( 0 \mid 1,0\right)\operatorname{P}_{1}\left( 0 \mid 1,1\right)+2\operatorname{P}_{1}\left( 1 \mid 0,0\right)\operatorname{P}_{1}\left( 0 \mid 1,1\right)}.
\end{align*}
At equilibrium, the probability that the agents pick the same alternatives (i.e., coordinate) is
\[
\mu_{(1,1)}+\mu_{(0,0)}=\dfrac{1}
{1+\dfrac{2}{\dfrac{\operatorname{P}_{1}\left( 1 \mid 0,1\right)}{\operatorname{P}_{1}\left( 0 \mid 1,1\right)}+\dfrac{\operatorname{P}_{1}\left( 0 \mid 1,0\right)}{\operatorname{P}_{1}\left( 1 \mid 0,0\right)}}}.
\]
Let us also suppose that each agent pays attention to the other agent if (and only if) their choices coincide. That is, $\operatorname{Q}\left(0,0\right)=\operatorname{Q}\left(1,1\right) =1$ and $\operatorname{Q}\left(0,1\right)=\operatorname{Q}\left(1,0\right) =0$. Then the probability that agents pick the same alternatives simplifies to
\[
\operatorname{Pr}_{\text{same}}=\dfrac{1}
{1+\dfrac{2}{\dfrac{\operatorname{R}\left( 1 \mid 0,0\right)}{\operatorname{R}\left( 0 \mid 0,1\right)}+\dfrac{\operatorname{R}\left( 0 \mid 0,0\right)}{\operatorname{R}\left( 1 \mid 1,0\right)}}}.
\]
If, in contrast, agents select each other only when their choices are different (i.e., $\operatorname{Q}\left(0,0\right)=\operatorname{Q}\left(1,1\right) =0$ and $\operatorname{Q}\left(0,1\right)=\operatorname{Q}\left(1,0\right) =1$), then 
\[
\operatorname{Pr}_{\text{diff}}=\dfrac{1}
{1+\dfrac{2}{\dfrac{\operatorname{R}\left( 1 \mid 0,1\right)}{\operatorname{R}\left( 0 \mid 0,0\right)}+\dfrac{\operatorname{R}\left( 0 \mid 1,0\right)}{\operatorname{R}\left( 1 \mid 0,0\right)}}}.
\]
Suppose that agents are indifferent between alternatives if the peer is not selected (i.e., $\operatorname{R}(1\mid 0, 0)=0.5$). Then, since $f(x)=x(1-x)$ achieves its maximum at 0.5, we have that
\begin{align*}
\operatorname{R}\left( 1 \mid 0,0\right)\operatorname{R}\left( 0 \mid 0,0\right)&=\operatorname{R}\left( 1 \mid 0,0\right)(1-\operatorname{R}\left( 1 \mid 0,0\right))\\
&\geq \max\{\operatorname{R}\left( 1 \mid 0,1\right)(1-\operatorname{R}\left( 1 \mid 0,1\right)), \operatorname{R}\left( 1 \mid 1,0\right)(1-\operatorname{R}\left( 1 \mid 1,0\right))\}\\
&=\max\{\operatorname{R}\left( 1 \mid 0,1\right)\operatorname{R}\left( 0 \mid 0,1\right), \operatorname{R}\left( 1 \mid 1,0\right)\operatorname{R}\left( 0 \mid 1,0\right)\}.
\end{align*}
Thus, 
\[
\dfrac{\operatorname{R}\left( 1 \mid 0,0\right)}{\operatorname{R}\left( 0 \mid 0,1\right)}> \dfrac{\operatorname{R}\left( 1 \mid 0,1\right)}{\operatorname{R}\left( 0 \mid 0,0\right)}\text{ and } \dfrac{\operatorname{R}\left( 0 \mid 0,0\right)}{\operatorname{R}\left( 1 \mid 1,0\right)}> \dfrac{\operatorname{R}\left( 0 \mid 1,0\right)}{\operatorname{R}\left( 1 \mid 0,0\right)}
\]
and 
\[
\operatorname{Pr}_{\text{same}}>\operatorname{Pr}_{\text{diff}}.
\]
We can interpret this result by saying that when the peer selection is based on choice similarity, then agents with similar preferences select each other more often, and the equilibrium frequency by which they select the same alternatives is higher. 

Let us finally compare these results with the standard model where agents always pay attention to all their peers. In this case, we have $\operatorname{P}_{1}\left( 1 \mid 1,1\right) = \operatorname{P}_{1}\left( 1 \mid 0,1\right) = \operatorname{R}\left( 1 \mid 0,1\right)$ and $\operatorname{P}_{1}\left( 1 \mid 0,0\right) = \operatorname{P}_{1}\left( 1 \mid 1,0\right) = \operatorname{R}\left( 1 \mid 1,0\right)$. Thus, the probability that agents pick the same alternatives is
\[
\operatorname{Pr}_{\text{std}}=\dfrac{1}
{1+\dfrac{2}{\dfrac{\operatorname{R}\left( 1 \mid 0,1\right)}{\operatorname{R}\left( 0 \mid 0,1\right)}+\dfrac{\operatorname{R}\left( 0 \mid 1,0\right)}{\operatorname{R}\left( 1 \mid 1,0\right)}}}.
\]

Interestingly, if the peer effects on preferences are positive (i.e., $\operatorname{R}\left( 1 \mid 0,1\right)>\operatorname{R}\left( 1 \mid 0,0\right)>\operatorname{R}\left( 1 \mid 1,0\right)$), then 

\[
\dfrac{\operatorname{R}\left( 1 \mid 0,1\right)}{\operatorname{R}\left( 0 \mid 0,1\right)}> \dfrac{\operatorname{R}\left( 1 \mid 0,0\right)}{\operatorname{R}\left( 0 \mid 0,1\right)}\text{ and } \dfrac{\operatorname{R}\left( 0 \mid 1,0\right)}{\operatorname{R}\left( 1 \mid 1,0\right)}> \dfrac{\operatorname{R}\left( 0 \mid 0,0\right)}{\operatorname{R}\left( 1 \mid 1,0\right)}
\]
and
\[
\operatorname{Pr}_{\text{std}} > \operatorname{Pr}_{\text{same}}>\operatorname{Pr}_{\text{diff}}.
\]
However, if the peer effects are negative (i.e., $\operatorname{R}\left( 1 \mid 0,1\right)<\operatorname{R}\left( 1 \mid 0,0\right)<\operatorname{R}\left( 1 \mid 1,0\right)$), then 

\[
\dfrac{\operatorname{R}\left( 1 \mid 0,1\right)}{\operatorname{R}\left( 0 \mid 0,1\right)}< \dfrac{\operatorname{R}\left( 1 \mid 0,1\right)}{\operatorname{R}\left( 0 \mid 0,0\right)}\text{ and } \dfrac{\operatorname{R}\left( 0 \mid 1,0\right)}{\operatorname{R}\left( 1 \mid 1,0\right)}< \dfrac{\operatorname{R}\left( 0 \mid 1,0\right)}{\operatorname{R}\left( 1 \mid 0,0\right)}
\]
and
\[
\operatorname{Pr}_{\text{same}}>\operatorname{Pr}_{\text{diff}}>\operatorname{Pr}_{\text{std}}.
\]
\hfill $\square$
\end{example}

\section{Identification of the Model}\label{sec: identification}

\subsection{Identification of the Model from the CCPs}\label{sec: modelid}

\noindent This section shows that all parts of the model can be recovered from a long sequence of choices. These parts include the network structure, the random preferences and the peer selection mechanism. That is, we will recover
\[
\left(\mathcal{N}_a\right)_{a\in\mathcal{A}}, \left(\operatorname{R}_a\right)_{a\in\mathcal{A}}, \left(\operatorname{Q}_a\right)_{a\in\mathcal{A}}.
\]
\noindent The starting point of our identification argument is the set of CCPs $\operatorname{P} = \left(\operatorname{P}_a\right)_{a\in\mathcal{A}}$ that can be calculated from the long sequence of choices made by the group members. Recall that, for each $v \in \mathcal{Y}$ and $a \in \mathcal{A}$, we have that
\begin{equation*}
\operatorname{P}_{a}\left( v \mid \mathbf{y}\right) =\sum\nolimits_{\mathcal{N} \subseteq \mathcal{N}_{a}}\operatorname{R}^{a}\left(v \mid \mathbf{y}, \mathcal{N}\right)\operatorname{S}^{a}\left(\mathcal{N} \mid \mathbf{y}, \mathcal{N}_a \right)
\end{equation*}
\noindent where
\begin{equation*}
\operatorname{R}^a(v\mid\mathbf{y},\mathcal{N})=\operatorname{R}_a(v\mid y_a,\operatorname{N}_a(\mathbf{y},\mathcal{N})) 
\end{equation*}
and 
\begin{equation*}
\operatorname{S}^{a}\left(\mathcal{N} \mid \mathbf{y}, \mathcal{N}_a \right) = \prod_{a'\in\mathcal{N}}\operatorname{Q}_{a}(a'\mid y_a,y_{a'}) \prod_{a'\in\mathcal{N}_a\setminus\mathcal{N}}(1-\operatorname{Q}_{a}(a'\mid y_a,y_{a'})).
\end{equation*}

\begin{proposition}\label{prop: network}
    Under Assumptions~\ref{ass: MM},~\ref{ass: simple}, and ~\ref{ass: regularity}, $\mathcal{N}_a$ is identified from $\operatorname{P}_a$ for every $a$.
\end{proposition}

\setcounter{aux}{\value{example}}
\setcounter{example}{\value{eg1}}
\begin{example}[continued]
Let us consider Agent 2. (The identification strategy is similar for the other agents.) The probability that Agent 2 selects alternative 1 given choice configuration $\mathbf{y}$ is
\begin{equation*}
\operatorname{P}_{2}\left( 1 \mid \mathbf{y}\right) =\operatorname{R}_{2}(1\mid y_2,\Char{y_{1} = 0},\Char{y_{1} = 1},\Char{y_{1} = 2})\operatorname{Q}_{2}(1\mid y_2,y_1) + \operatorname{R}_{2}(1\mid y_2,0,0,0)(1-\operatorname{Q}_{2}(1\mid y_2,y_1)).
\end{equation*}

\noindent Take, to simplify the exposition, $\mathbf{y}=\mathbf{0}$. After some manipulation, the changes in $\operatorname{P}_{2}$ when we change the choices of Agents 1, 3, and 4 from 0 to 1 are, respectively,
\begin{align*}
\operatorname{P}_{2}\left( 1 \mid 1,0,0,0\right) - \operatorname{P}_{2}\left( 1 \mid 0,0,0,0\right) &= (\operatorname{R}_2(1\mid 0,0,1,0) - \operatorname{R}_2(1\mid 0,0,0,0))\operatorname{Q}_{2}(1\mid 0,1)
\\
&- (\operatorname{R}_2(1\mid 0,1,0,0)-\operatorname{R}_2(1\mid 0,0,0,0))\operatorname{Q}_{2}(1\mid 0,0),
\\
\operatorname{P}_{2}\left( 1 \mid 0,0,1,0\right) - \operatorname{P}_{2}\left( 1 \mid 0,0,0,0\right) &= 0,
\\
\operatorname{P}_{2}\left( 1 \mid 0,0,0,1\right) - \operatorname{P}_{2}\left( 1 \mid 0,0,0,0\right) &= 0.
\end{align*}
\noindent Proposition~\ref{prop: network} identifies the peers of Agent 2 as those for which these changes differ from 0. Assumption~\ref{ass: regularity} guarantees that the two terms in the right-hand side of the first line are not of equal size. Following this idea, we correctly recover $\mathcal{N}_2 = \left\{1\right\}$.
\hfill $\square$
\end{example}

To recover the choice rules and the peer selection mechanisms, we implement two variants of an idea that we describe next by example. The idea relies on manipulating the size of the set $\mathcal{N}_a$.

\setcounter{aux}{\value{example}}
\setcounter{example}{\value{eg1}}
\begin{example}[continued]
Recall that (by Proposition~\ref{prop: network}) we can recover the peer group of each agent. Thus, for instance, we can learn that 
$
\mathcal{N}_{1}=\left\{2, 3\right\}
$.
Consider the following thought experiment. Take Agent 1 and assume that we can $\emph{remove}$ agents from her reference group and recover her CCPs. In doing so, we will vary $\abs{\mathcal{N}_1}$ along 0, 1, and 2. The probabilities by which Agent 1 select alternative $v$ if we remove Agents 2 and 3, only Agent 3, only Agent 2, and none of them from her reference group are, respectively, are given by
{\small
\begin{align*}
\operatorname{P}_{1}\left( v \mid \mathbf{y}\right) =&\operatorname{R}_1(v\mid y_1,0,0,0)
\\
\operatorname{P}_{1}\left( v \mid \mathbf{y}\right) =&\operatorname{R}_1(v\mid y_1,\Char{y_{2} = 0},\Char{y_{2} = 1},\Char{y_{2} = 2})\operatorname{Q}_1(2 \mid y_1,y_2) + \operatorname{R}_1(v\mid y_1,0,0,0)(1-\operatorname{Q}_1(2 \mid y_1,y_2))
\\
\operatorname{P}_{1}\left( v \mid \mathbf{y}\right) =&\operatorname{R}_1(v\mid y_1,\Char{y_{3} = 0},\Char{y_{3} = 1},\Char{y_{3} = 2})\operatorname{Q}_1(3 \mid y_1,y_3) + \operatorname{R}(v\mid y_1,0,0,0)(1-\operatorname{Q}_1(3 \mid y_1,y_3))
\\
\operatorname{P}_{1}\left( v \mid \mathbf{y}\right) = &\operatorname{R}(v\mid y_1,\sum_{a'=2,3}\Char{y_{a'} = 0},\sum_{a'=2,3}\Char{y_{a'} = 1},\sum_{a'=2,3}\Char{y_{a'} = 2})\operatorname{Q}_1(2 \mid y_1,y_2) \operatorname{Q}_1(3 \mid y_1,y_3)
\\
&+ \operatorname{R}_1(v\mid y_1,\Char{y_2 = 0},\Char{y_2 = 1},\Char{y_2 = 2})\operatorname{Q}_1(2 \mid y_1,y_2) (1-\operatorname{Q}_1(3 \mid y_1,y_3))
\\
&+ \operatorname{R}_1(v\mid y_1,\Char{y_3 = 0},\Char{y_3 = 1},\Char{y_3 = 2})(1-\operatorname{Q}_1(2 \mid y_1,y_2))\operatorname{Q}_1(2 \mid y_1,y_3)
\\
&+ \operatorname{R}_1(v\mid y_1,0,0,0)(1-\operatorname{Q}_1(2 \mid y_1,y_2))(1-\operatorname{Q}_1(3 \mid y_1,y_3)).
\end{align*}
}
The above system relates four observed CCPs and six unknowns (i.e., the choice rule and the peer selection probabilities). It allows the researcher to uniquely recover one of the unknowns ($\operatorname{R}_1(v\mid y_1,0,0,0)$) and partially identify the other ones.  
\hfill $\square$
\end{example}
The example shows that if the researcher $\emph{could}$ manipulate the reference groups (by removing some links) the observed data would be informative about both the choice rules and the peer selection mechanism. In the applications we have in mind, however, the researcher has no control over reference groups and cannot arbitrarily modify them. To address this issue, we introduce two models in which the required variation arises endogenously through different channels. In each case, we also add restrictions to restore full (rather than partial) identification. 

\subsubsection*{Different Sizes of Reference Groups across Agents} 
In the first setup, we assume that agents can be partitioned into a finite set of types $\mathcal{H}=\{1, 2, \dots, H\}$, $H\leq A$. The type of each agent is given by a known mapping $h:\mathcal{A}\to\mathcal{H}$ where $h(a)$ encodes the type of Agent $a$. These types might be related to covariates or other individual characteristics that the researcher observes. Each type incorporates two dimensions: a random choice rule that captures the preferences of the agent over the set of alternatives; and a limited attention mechanism that connects the preferences of each agent to the choices of her active peers. These types do not impose any restrictions on the reference groups of peers. That is, two agents of the same type might have very different reference groups.
\setcounter{aux}{\value{example}}
\setcounter{example}{\value{eg1}}
\begin{example}[continued]
Suppose that Agents $1$ and $2$ have a college degree while Agents $3$ and $4$ do not. Thus, we can define $h$ as $h(1)=h(2)=1$ and $h(3)=h(4)=2$ with $\mathcal{H}=\{1,2\}$.
\hfill $\square$
\end{example}

The main assumption is as follows.

\begin{assumption}\label{ass: types}
    For each $a$, $\mathbf{y}$, $a'\in\mathcal{N}_a$, $\mathcal{N}\subseteq\mathcal{N}_a$, and $v$
    \begin{enumerate}
        \item $\operatorname{Q}_a(a'\mid y_a,y_{a'})= \operatorname{Q}_{h(a)}(y_a,y_{a'})$; and
        \item $\operatorname{R}_a(v\mid y_a,\operatorname{N}_a(\mathbf{y},\mathcal{N}))=\operatorname{R}_{h(a)}(v\mid y_a,\operatorname{N}_a(\mathbf{y},\mathcal{N}))$.
    \end{enumerate}
\end{assumption}
Assumption~\ref{ass: types} imposes that agents belonging to the same type share identical functional forms for both selection probabilities and choice rules. That is, if two agents have the same type, they use the same peer-selection rule to determine which peers are considered and the same choice rule to choose among alternatives, up to observable state variables. Types therefore represent clusters of agents that behave similarly in how they form attention over peers and how they translate the set of considered peers into choices.

In our empirical application, an agent is defined as a firm-market pair. The type of an agent is the firm to which it belongs. The coefficients governing how competitor (peer) presence affects attention and profitability are firm specific but do not vary across markets. Markets differ through observable characteristics and the realized competitive environment, but the mapping from these variables into attention and choice probabilities is common within firm. Thus, types partition agents by firm, and each firm defines a behavioral cluster with its own selection and payoff parameters.

The identification result we describe below requires variation in the sizes of reference groups among agents within each type. As before, this type of variation leads to partial identification of the choice rules and peer selection mechanisms. Under Assumptions~\ref{ass: MM} and~\ref{ass: simple}, the CCPs of Agent $a$ can be presented as a binomial mixture (or convolution) model. For each $v \in \mathcal{Y}$ and $a \in \mathcal{A}$, $\operatorname{P}_{a}\left( v \mid \mathbf{y}\right)$ can be written as
\begin{equation*}
\sum_{n_0=0}^{\operatorname{N}^0_{a}\left(\mathbf{y}, \mathcal{N}_{a}\right)}\dots\sum_{n_Y=0}^{\operatorname{N}^Y_{a}\left(\mathbf{y}, \mathcal{N}_{a}\right)}\operatorname{R}_{h(a)}(v\mid y_a,n_0,\dots,n_Y) \prod_{y=0}^{Y}\binom{\operatorname{N}^y_{a}\left(\mathbf{y}, \mathcal{N}_{a}\right)}{n_y}\operatorname{Q}_{h(a)}(y_a,y)^{n_y}\left(1-\operatorname{Q}_{h(a)}(y_a,y)\right)^{\operatorname{N}^Y_{a}\left(\mathbf{y}, \mathcal{N}_{a}\right)-n_y},
\end{equation*}
where $\sum_{y=0}^{Y}\operatorname{N}^y_{a}\left(\mathbf{y}, \mathcal{N}_{a}\right)=\abs{\mathcal{N}_a}$ for all $\mathbf{y} \in \mathcal{Y}^A$. After learning $\left(\mathcal{N}_a\right)_{a\in\mathcal{A}}$ we recover $ \left(\operatorname{R}_a\right)_{a\in\mathcal{A}}, \left(\operatorname{Q}_a\right)_{a\in\mathcal{A}}$ taking advantage of variation on the size of the reference groups, i.e. $\abs{\mathcal{N}_{a}}$. 

To obtain full identification, we invoke two other restrictions. In the next assumption, let $\mathbf{0}^{\abs{\mathcal{N}}}_v$ be a size $\abs{\mathcal{Y}}$ vector of aggregate choices where all active peers select option $v$. (That is, a vector of 0s except for the $v$-component that is $\abs{\mathcal{N}}$.)

\begin{assumption}\label{ass: average}
    For each $a$, $\mathbf{y}$, $\mathcal{N}\subseteq\mathcal{N}_a$,  $\mathcal{N}\neq\emptyset$, $v$, and $v'$
    \begin{enumerate}
        \item $\operatorname{R}_a\left(v\mid y_a,\mathbf{0}^{1}_{v'}\right)=\operatorname{R}_a\left(v\mid y_a,\mathbf{0}^{\abs{\mathcal{N}}}_{v'}\right)$; and
        \item $\operatorname{R}_a\left(v\mid y_a,\operatorname{N}_a(\mathbf{y},\mathcal{N})\right)\neq \operatorname{R}_a\left(v\mid y_a,\mathbf{0}\right)$.
    \end{enumerate}
\end{assumption}

Assumption~\ref{ass: average}(i) states that the choice rule applied when all active peers select the same alternative does not change with the number of active peers. This restriction is satisfied if, for instance, the choice rule depends on the proportion of peers selecting each alternative ---an assumption that is rather standard in the peer effects literature. Assumption~\ref{ass: average}(ii) simply states that the choice rule of the empty active peer set differs from the choice rule of the non-empty set for any choice configuration.

Proposition~\ref{prop: network} states that we can recover the reference group of each agent from the CCPs. We next use a recursive argument to show that in this setting we can recover the choice rules and the peer selection mechanism if we have variation on the number of peers in the reference groups of agents of the same type. Specifically, let 
\begin{equation*}
\operatorname{N}_t \equiv \left\{\abs{\mathcal{N}_a}\::\:a\in\mathcal{A},\:h(a)=t\right\}
\end{equation*}
be the set of all $\emph{sizes}$ of reference groups for agents of type $t$. We next show that having at least three agents of the same type with different numbers of potential peers provides enough variation to identify peer selection probabilities and choice rules with no or just one active peer.
\begin{proposition}\label{prop: selection}
     In addition to Assumptions~\ref{ass: MM},~\ref{ass: simple},~\ref{ass: regularity},~\ref{ass: types}, and~\ref{ass: average}, suppose that $\abs{\operatorname{N}_t}\geq 3$ for each $t\in\mathcal{H}$. Then, $\operatorname{Q}^a$, $\operatorname{R}^a(\cdot|\cdot,\emptyset)$ and $\operatorname{R}^a(\cdot|\cdot,\{a'\})$, $a'\in\mathcal{N}_a$, are identified from $\operatorname{P}_a$ for every $a$.
\end{proposition}
Although we may not identify the choice rule for all possible active peer groups nonparametrically, knowing it for the empty set and for singletons is enough if we further assume preferences are represented by the standard linear-in-means multinomial logit model. Below, assuming richer variation in the sizes of the reference groups, we show that random choice rules can be nonparametrically identified without restricting attention to the logit model.

\begin{proposition}\label{prop: logit rule}
    If the choice rule satisfies
    \[
    \operatorname{R}^a(v\mid\mathbf{y},\mathcal{N})=\dfrac{e^{\alpha_{h(a),v}(y_a)+\beta_{h(a),v}(y_a)\operatorname{N}^v_{a}(\mathbf{y},\mathcal{N})/\abs{\mathcal{N}}}}{\sum_{v'\in\mathcal{Y}}e^{\alpha_{h(a),v'}(y_a)+\beta_{h(a),v'}(y_a)\operatorname{N}^{v'}_{a}(\mathbf{y},\mathcal{N})/\abs{\mathcal{N}}}}
    \]
    with the normalization $\alpha_{h(a),0}(y_a)=0$, then $\operatorname{R}^a$ is identified from $\operatorname{R}^a(\cdot|\cdot,\emptyset)$ and $\operatorname{R}^a(\cdot|\cdot,\{a'\})$, $a'\in\mathcal{N}_a$.
\end{proposition}

Once the peer selection mechanism and the choice rule (for the cases of no and just one active peer) are recovered, we can recursively identify the choice rule for all other subsets of active peers if we have full variation in the size of connected agents within each type. For example, suppose Agent $a$ has 2 peers, $a_1$ and $a_2$ (i.e., $\mathcal{N}_a=\{a_1,a_2\}$). Then
\begin{align*}
\operatorname{P}_a(v\mid\mathbf{y})&=\operatorname{S}^a(\emptyset\mid\mathbf{y},\mathcal{N}_a)\operatorname{R}^a(v\mid\mathbf{y},\emptyset)+\operatorname{S}^a(\{a_1\}\mid\mathbf{y},\mathcal{N}_a)\operatorname{R}^a(v\mid\mathbf{y},\{a_1\})\\
&+\operatorname{S}^a(\{a_2\}\mid\mathbf{y},\mathcal{N}_a)\operatorname{R}^a(v\mid\mathbf{y},\{a_2\})
+\operatorname{S}^a(\{a_1,a_2\}\mid\mathbf{y},\mathcal{N}_a)\operatorname{R}^a(v\mid\mathbf{y},\{a_1,a_2\}).
\end{align*}
Hence, since the selection mechanism and the choice rule with no and just one active peer are identified, we can recover the choice rule with two active peers, $\operatorname{R}^a(v\mid\mathbf{y},\{a_1,a_2\})$. Repeating the same argument for someone who has three peers, we can identify the choice rule with three active peers. Thus, with enough variation in the sizes of the reference groups for different agents within each type, we can identify $\operatorname{R}^a$ without any parametric restrictions.

\begin{assumption}\label{ass: full variation}
    For every type $t\in\mathcal{H}$, 
    \[
    \{2,3,\dots,\max_{\{a\in\mathcal{A},\:h(a)=t\}}\abs{\mathcal{N}_a}\}\subseteq\operatorname{N}_t.
    \]
\end{assumption}
\noindent Assumption~\ref{ass: full variation} means that within each type there is some agent with two peers, someone with three peers, etc. This assumption is similar to the full menu variation (of alternatives) in the stochastic choice literature \citep{aguiar2023random}. 

The proof of the following result follows from Proposition~\ref{prop: selection} and the above discussion.
\begin{proposition}\label{prop: recursive}
    Suppose the assumptions of Proposition~\ref{prop: selection} are satisfied. If, moreover, Assumption~\ref{ass: full variation} is satisfied, then  $\operatorname{Q}^a$ and $\operatorname{R}^a$ are identified from $\operatorname{P}_a$ for every $a$. 
\end{proposition}

\setcounter{aux}{\value{example}}
\setcounter{example}{\value{eg1}}
\begin{example}[continued]
\noindent Assume the four agents in the example are of the same type. Thus, to simplify the exposition, we will drop the subindices in $Q$ and $R$. Recall that, by Proposition~~\ref{prop: network}, we can recover the peer group of each of them. Thus, we can learn that 
\[
\mathcal{N}_{1}=\left\{2, 3\right\},\quad \mathcal{N}_{2}=\left\{1\right\},\quad \mathcal{N}_{3}=\left\{2\right\}, \quad\mathcal{N}_{4}=\emptyset.
\]
After some manipulation, we get that
\begin{equation*}
\frac{\operatorname{P}_1(v\mid v',v'',v'',v'')-\operatorname{P}_4(v\mid v'',v'',v'',v')}{\operatorname{P}_2(v\mid v'',v',v'',v'')-\operatorname{P}_4(v\mid v'',v'',v'',v')} = 2 - \operatorname{Q}(v',v'').
\end{equation*}
\noindent Since the left-hand side is observed, we can recover $\operatorname{Q}(v',v'')$ for each pair $v',v'' \in \mathcal{Y}$. That is, the peer selection mechanism can be recovered by combining (in a specific way) the CCPs of agents that differ regarding number of peers in their reference groups. We next show how to recursively recover the choice rule.  

In the example, Agent 4 has no peers. It follows that 
\begin{equation*}
\operatorname{P}_{4}\left( v \mid \mathbf{y}\right) =\operatorname{R}(v\mid y_4,\mathbf{0}).
\end{equation*}
\noindent This means that we can recover from the CCPs of Agent 4 the choice rules for the empty set of active peers. Let $\mathbf{0}_{v'}^1$ be a vector of three elements, all of which are 0 except for the one corresponding to (the different) alternative $v'$ which is 1. Let us consider Agent 2
\begin{equation*}
\operatorname{P}_{2}\left( v \mid y_1=v',y_2,y_3,y_4\right) = \operatorname{R}(v\mid y_2,\mathbf{0}_{v'}^1)\operatorname{Q}(y_2,v') + \operatorname{R}(v\mid y_2,\mathbf{0})(1-\operatorname{Q}(y_2,v')).
\end{equation*}
\noindent Since we already identified two of the three terms on the right-hand side of the last expression, we can recover the choice rule for one active peer from the CCPs of Agent 2. Let $\mathbf{0}_{v',v''}^1$ be a vector of three elements, all of which are 0 except for the ones corresponding to alternatives $v'$ and $v''$ which are 1. Let us next consider Agent 1
{\small
\begin{align*}
\operatorname{P}_{1}\left( v \mid y_1,y_2=v',y_3=v'',y_4\right) &= \operatorname{R}(v\mid y_1,\mathbf{0}_{v',v''}^1)\operatorname{Q}(y_1,v')\operatorname{Q}(y_1,v'') + \operatorname{R}(v\mid y_1,\mathbf{0})(1-\operatorname{Q}(y_1,v'))(1-\operatorname{Q}(y_1,v'')) 
\\
&+ \operatorname{R}(v\mid y_1,\mathbf{0}_{v'}^1)\operatorname{Q}(y_1,v')(1-\operatorname{Q}(y_1,v'')) + \operatorname{R}(v\mid y_1,\mathbf{0}_{v''}^1)(1-\operatorname{Q}(y_1,v'))\operatorname{Q}(y_1,v''). 
\end{align*}}
Since we already identified five of the six terms on the right-hand side of the last expression, we can recover the choice rule for two active peers (that have chosen different alternatives) from the CCPs of Agent 1. 
\hfill $\square$
\end{example}

\subsubsection*{Outside Option with no Peer Effects} In the second setting, we assume that the choice rules are not affected by active peers selecting the outside option 0. For the next assumption, let $\operatorname{N}=\left(\operatorname{N}^v\right)_{v\in\mathcal{Y}}$ indicate a vector of the number of peers that select each alternative.

\begin{assumption}\label{ass: outside}
    For each $a$, $\mathbf{y}$, and $v\neq 0$, (i) the choice rule $\operatorname{R}_a\left(v\mid y_a,\operatorname{N}\right)$ does not vary with $\operatorname{N}^{0}$; and  (ii) $\operatorname{R}_{a}\left(v \mid \ y_a,\mathbf{0}^1_{v}\right)-\operatorname{R}_{a}\left(v \mid \ y_a,\mathbf{0}\right)\neq 0$ for all $v\in\mathcal{Y}\setminus\{0\}$.
\end{assumption}

Assumption~\ref{ass: outside}(i) allows us to mimic variation in reference groups for different choice configurations at the level of each agent. As we described before, this variation produces useful information to recover the model. In our case, it leads to a partially identified model where the choice rule and peer-selection mechanism can be recovered up to one element. Assumption~\ref{ass: outside}(ii) just means that there is a peer effect in preferences. 

The next assumption offers a condition that suffices to recover all elements of the model. 

\begin{assumption}\label{ass: always}
    For each $a \in \mathcal{A}$ with $\abs{\mathcal{N}_a} \geq 1$, and all $v_1,v_2\in\mathcal{Y}$, the value of either
    \[
\min_{a'\in\mathcal{N}_a}\operatorname{Q}_{a}(a'\mid v_1,v_2)\quad
    \text{ or }\quad
    \max_{a'\in\mathcal{N}_a}\operatorname{Q}_{a}(a'\mid v_1,v_2)
    \]
    is known.
\end{assumption}
Assumption~\ref{ass: always} requires the existence of a known maximal or minimal selection probability for all possible configurations of choices that Agent $a$ and the target Agent $a'$ can take. It is satisfied if for every configuration of $v_1$ and $v_2$ there is a peer that is always selected: $\operatorname{Q}_{a}(a'\mid v_1,v_2)=1$ for some $a'$. (This peer does not have to be the same for all possible $v_1$ and $v_2$.)  The choice of this peer works as an exclusion restriction by shifting the choice rule without affecting the selection process. Importantly, the set of peers where the maximum or minimum in Assumption~\ref{ass: always} is achieved does not need to be known. For example, if we assume that one peer is always selected, this peer does not need to be known and can be identified from the data.

\begin{proposition}\label{prop: idoutside}
    If Assumptions~\ref{ass: MM},~\ref{ass: simple}, ~\ref{ass: regularity}, ~\ref{ass: outside}, and ~\ref{ass: always} are satisfied, then  $\operatorname{Q}^a$ and $\operatorname{R}^a$ are identified from $\operatorname{P}_a$ for every $a$. 
\end{proposition}

\setcounter{aux}{\value{example}}
\setcounter{example}{\value{eg1}}
\begin{example}[continued]
\noindent Assume that all agents are of the same type. Recall that, by Proposition~~\ref{prop: network}, we know
\[
\mathcal{N}_{1}=\left\{2, 3\right\}.
\]
Take Agent 1. If Assumption~~\ref{ass: outside} holds we can mimic variation in the reference groups by considering choice configurations in which some of the agents in her reference group select the outside option. Doing so, we obtain
\begin{align*}
&\operatorname{P}_{1}\left( v \mid y_1,0,0,0\right) =\operatorname{R}(v\mid y_1,0,0,0)
\\
&\operatorname{P}_{1}\left( v \mid y_1,y_2,0,0\right) =\operatorname{R}(v\mid y_1,\Char{y_{2} = 0},\Char{y_{2} = 1},\Char{y_{2} = 2})\operatorname{Q}(y_1,y_2) + \operatorname{R}(v\mid y_1,0,0,0)(1-\operatorname{Q}(y_1,y_2))
\\
&\operatorname{P}_{1}\left( v \mid y_1,0,y_3,0\right) =\operatorname{R}(v\mid y_1,\Char{y_{3} = 0},\Char{y_{3} = 1},\Char{y_{3} = 2})\operatorname{Q}(y_1,y_3) + \operatorname{R}(v\mid y_1,0,0,0)(1-\operatorname{Q}(y_1,y_3))
\\
&\operatorname{P}_{1}\left( v \mid y_1,y_2,y_3,0\right) = \operatorname{R}(v\mid y_1,\sum_{a'=2,3}\Char{y_{a'} = 0},\sum_{a'=2,3}\Char{y_{a'} = 1},\sum_{a'=2,3}\Char{y_{a'} = 2})\operatorname{Q}(y_1,y_2) \operatorname{Q}(y_1,y_3)
\\
&\qquad \qquad \quad + \operatorname{R}(v\mid y_1,\Char{y_2 = 0},\Char{y_2 = 1},\Char{y_2 = 2})\operatorname{Q}(y_1,y_2) (1-\operatorname{Q}(y_1,y_3))
\\
&\qquad \qquad \quad + \operatorname{R}(v\mid y_1,\Char{y_3 = 0},\Char{y_3 = 1},\Char{y_3 = 2})(1-\operatorname{Q}(y_1,y_2))\operatorname{Q}(y_1,y_3)
\\
&\qquad \qquad \quad + \operatorname{R}(v\mid y_1,0,0,0)(1-\operatorname{Q}(y_1,y_2))(1-\operatorname{Q}(y_1,y_3)).
\end{align*}
\noindent As in the thought experiment we described earlier, this approach generates a system of four observed probabilities that depend on six unknowns related to the choice rules and peer selection probabilities. The system allows the researcher to uniquely recover one of the unknowns and partially identify the other ones. If we also assume that either Agent 2 or 3 is always an active peer, then we can uniquely recover the choice rule and the peer-selection mechanism. 
\hfill $\square$
\end{example}

The two identification strategies we offer are different in the sets of assumptions they are imposing. (Table~\ref{tab: strategies comp} summarizes them.) Most of these assumptions, are not mutually exclusive so these two strategies can be combined if a particular application requires it. Moreover, these identification results do not impose parametric functional form restrictions on peer-selection and choice rules. If one is willing to impose parametric restrictions, as we do in our empirical illustration, then some of these assumptions can be weakened or completely dropped (e.g., full support in networks sizes is not needed anymore). 
\begin{table}[h!]
\centering
\caption{Comparison of Identification Assumptions (Section 4.1)}
\begin{tabular}{lcc}
\hline
 & \textbf{Strategy 1} & \textbf{Strategy 2} \\
\hline
Types & Yes & No \\
Variation in $|\mathcal{N}_a|$ & Yes & No \\
Full support in network sizes & Yes  & No \\
Invariance when all peers choose same option & Yes & No \\
Outside option with no peer effects & No & Yes \\
Known max/min selection probability & No & Yes \\
\hline
\end{tabular}
\label{tab: strategies comp}
\end{table}

\subsection{Identification of Conditional Choice Probabilities}

\noindent This section shows that the CCPs, $\operatorname{P}$, and the rates of the Poisson alarm clocks, $\lambda$, can be recovered from a long sequence of choices. We assume the researcher observes the choices of the agents at time intervals of length $\Delta$ and can consistently estimate $\Pr\left(\mathbf{y}^{t+\Delta }=\mathbf{y}'\mid\mathbf{y}^{t}=\mathbf{y}\right)$ for each pair $\mathbf{y}',\mathbf{y}\in{\mathcal{Y}}^{A}$, to construct a matrix $\mathcal{P}\left( \Delta \right)$.\footnote{Here again, we assume that the choice configurations are ordered according to the lexicographic order when we construct $\mathcal{P}\left(\Delta \right)$.} Matrix $\mathcal{P}\left( \Delta \right)$ relates to the transition rate matrix $\mathcal{W}$ by $\mathcal{P}\left( \Delta \right) =e^{\left( \Delta \mathcal{W}\right) }$. (Here $e^{\left( \Delta \mathcal{W}\right)}$ is the matrix exponential of $\Delta \mathcal{W}$.)   Often, the researcher observes the precise moment at which agents revise strategies and the choices at that time. In other cases, the researcher simply observes the configuration of choices at fixed time intervals ---e.g., every Monday. \citet{kashaev2025peer} refer to these two cases as Dataset 1 and 2, respectively. Formally, in Dataset 1, the researcher can consistently estimate $\mathcal{W}$, and, in Dataset 2, the researcher can consistently estimate $\mathcal{P}\left(
\Delta \right)$. In the second case, the identification question is whether (or under what extra restrictions) we can uniquely recover $\mathcal{W}$ from $\mathcal{P}\left( \Delta \right)$. The identification problem in Dataset 1 is straightforward. \citet{kashaev2025peer}, using Theorem~1 in \citet{blevins2018identification}, offer a mild condition under which the transition rate matrix can be identified from Dataset 2.\footnote{It is also known that if the researcher were to observe the dynamic system at two different intervals $\Delta_{1}$ and $\Delta_{2}$ that are not multiples of each other (see, for example, \citealp{blevins2017identifying} and the literature therein).} 

\begin{proposition}\label{ID2} If Assumptions~\ref{ass: MM} and~\ref{ass: types} hold, then the CCPs $\operatorname{P}$ and the rates of the Poisson alarm clocks $(\lambda_a)_{a\in\mathcal{A}}$ are identified from Dataset 1. If, moreover, $\mathcal{W}$ has distinct eigenvalues that do not differ by an integer multiple of $2\pi i/\Delta $, where $i$ denotes the imaginary unit, then $\operatorname{P}$ and $(\lambda_a)_{a\in\mathcal{A}}$ are generically identified from Dataset 2.
\end{proposition}

The restriction on eigenvalues of $\mathcal{W}$ is a regularity condition that is generically satisfied.\footnote{See \citet{blevins2017identifying} for a discussion of this assumption.} The key element in proving the second statement in Proposition~\ref{ID2} is that the transition rate matrix in our model is rather parsimonious. Since, at each time, at most one person revises her selection with a nonzero probability, $\mathcal{W}$ has many zeros in known locations.

\section{Application}\label{sec: application}

\noindent We study the dynamic expansion and contraction decisions of five leading national fast-food franchising chains ---Wallace, KFC, McDonald's, Burger King, and Dicos--- operating across geographically distinct markets. The empirical setting consists of a panel of fast-food chains operating in Chinese cities over several decades. Firms decide whether to open or close restaurants in each market, taking into account local market conditions and the presence of competitors. A key feature of our setting is that firms can endogenously update their set of active competitors over time. The possibility that firms might ignore some competitors reflects information frictions, bounded rationality, and the fact that only a subset of rivals could be strategically relevant.

\subsection{Empirical Model}
\noindent We first describe the model of firm expansion and contraction decisions and then introduce the specifications for selection of competitors and firm payoffs. Similar to \citet{kashaev2025peer}, we define a market at the level of a prefecture-level city, which ranks below a province and above a county in China's administrative structure. Thus, markets are geographically isolated from each other. We collect all unknown primitives by $\theta$.

\paragraph{Choice Set, Agents, Types, and Peers} The empirical unit of observation is a firm--market pair \( (f,m) \), where \( f \in \mathcal{F} = \{1,\dots,F\} \) indexes firms and \( m \in \mathcal{M} = \{1,\dots,M\} \) indexes geographic markets. Every firm $f$ decides whether to open a restaurant ($v=2$), do nothing ($v=1$), or close a restaurant ($v=0$) in market $m$. We call a pair $a=(f,m)$ an agent ---knowing the firm and the market identifies the agent and vice versa. Thus, $\mathcal{A}=\mathcal{F}\times\mathcal{M}$ and $\mathcal{Y}=\{0,1,2\}$.\footnote{We abuse notation a bit since $\mathcal{A}$ was previously defined as a set of the form $\{1,2,\dots, A\}$. To be consistent with the initial notation, we can take any one-to-one mapping $\tilde{a}:\mathcal{F}\times\mathcal{M}\to\{1,2,\dots, \abs{\mathcal{F}\times\mathcal{M}}\}$ and define an agent as $a=\tilde{a}(f,m)$.} Given that in our data, in about 77\% days with any action (restaurant closure or opening), action happens in more than one city, we assume that the agents have perfectly synchronized clocks. This assumption does not affect the identification argument in Section~\ref{sec: modelid}

In line with our assumption about types, we assume that firm parameters do not vary across markets: preference and selection parameters  are firm-specific but common to all markets where the firm operates. That is, in our application $h(a)=h((f,m))=f$. Next, we assume that the set of firms perceived as (potential) competitors varies across markets for exogenous reasons (such as entry history or geography) and that the researcher does not directly observe which firms are included. This variation in the size of potential competitors (or reference groups) is used for identification, as in the theoretical model.

Since these fast-food restaurants are small and operate under franchise arrangements, we argue that the managers of these restaurants (i) make decisions independently across markets and (ii) do not always consider every competitor within the market. Managers might ignore the information about competitors for different reasons. For example, the number of competitors' restaurants may be small or the manager may not think that a particular restaurant is a substitute.  In addition, monitoring competitors is costly in terms of time, data collection, analysis costs, and managers might not be able to process all market information at all times.

At the moment of deciding whether to open or close a restaurant, the attention that firm $f$ pays to her competitors within market $m$ depends on her own and competitor's past choices in market $m$. Moreover, since the markets are locally isolated, we follow the literature \citep[e.g.,][]{arcidiacono2016estimation} and assume the marginal profit of firm $f$ in market $m$ from opening a new restaurant is only affected by both its own and its competitors' openings in market $m$. Formally, for any $a=(f,m)$,  $a'\not\in\mathcal{N}_{a}$ if $m'\neq m$.

\paragraph{Observable Characteristics} Every market $m$ at every moment of time $t$ is characterized by observed market characteristics $S_{mt}$ (e.g., GDP and population density) that include a constant. Let $N_{at}$ denote the number of restaurants of Agent $a$ (i.e., the number of restaurants of firm $f$ in market $m$). Also define $S_t=(S_{mt})_{m\in\mathcal{M}}$ and $N_{t}=(N_{at})_{a\in\mathcal{A}=\mathcal{F}\times\mathcal{M}}$.

\paragraph{Peer selection} In our application, we assume that every firm may or may not consider any other firm that could operate in the market. Given the numbers of restaurants each firm had in the market at time $t$, $N_t$, the probability that Agent $a=(f,m)$ considers Agent $a'=(f',m')$ at time $t$ is
\[
\operatorname{Q}_f(N_{at},N_{a't},S_{mt})=\Char{a'\in\mathcal{N}_{a}}\operatorname{F}_{\tilde\varepsilon}\left(\bar{\tilde{\pi}}_{f}(N_{at},N_{a't},S_{mt};\theta)\right),
\]
where $\operatorname{F}_{\tilde\varepsilon}$ is a known c.d.f.; $\theta$ is the vector of unknown parameters; and $\bar{\tilde{\pi}}_{f}(N_{at},N_{a't},S_{mt};\theta)$ is the mean attention index of firm $f$, which is known up to $\theta$. We allow the current number of restaurants of firms $f$ and $f'$ in market $m$ to affect the attention index of firm $f$.

\paragraph{Payoff from  Opening or Closing a Restaurant} Conditional on a set of competitors being considered, the firm decides whether to open or close at least one new restaurant, or do nothing in that market based on its marginal profit $\pi_{at}$. This marginal profit captures not just the instantaneous (one period) profitability of an extra restaurant, but the expected profitability of the restaurant in the long run ---see the discussion in Section~\ref{sec: rc}. 

This is an ordered response model, with the choice being 
\[
v=\Char{\bar{\pi}_{f}\left(N_{at},\sum_{a'\in\mathcal{N}}N_{a't},S_{mt};\theta\right)+\varepsilon\geq \gamma_{1f}}+\Char{\bar{\pi}_{f}\left(N_{at},\sum_{a'\in\mathcal{N}}N_{a't},S_{mt};\theta\right)+\varepsilon\geq \gamma_{1f}+\gamma_{2f}},
\]
where $\varepsilon$ is an idiosyncratic random shock with a known  c.d.f. $\operatorname{F}_{\varepsilon}$;  $\bar{\pi}_{f}\left(N_{at},\sum_{a'\in\mathcal{N}}N_{a't},S_{mt};\theta\right)$ is the mean marginal profit of firm $f$ in market $m$ that depends on the own number of restaurants, the aggregate number of restaurants of considered competitors, and market characteristics. The profit function is known up to $\theta$. The threshold parameters $\gamma_{1f}$ and $\gamma_{2f}>0$ (in $\theta$) are unknown. 
\begin{rem} 
Formally, our application is using \emph{all} previous choices of firms as determinants of selection and marginal profits. That is, we have infinite history dependence. However, similar to \citet{kashaev2025peer}, since the model we estimate in our application is fully parametric, this should not lead to any issues.
\end{rem}

\paragraph{Model Implied CCP} Altogether, the above parametrization leads to a model implied distribution over three ordered outcomes for every firm in every market:
\begin{equation*}
\operatorname{Pr}_a(v\mid (N_{a't})_{a'\in \mathcal{N}_a\cup \{a\}},S_{mt};\theta),
\end{equation*}
which completely characterizes the probability of observing a new restaurant or a restaurant closure in a given market by a given firm conditional on the history and the market characteristics. When evaluated at the true parameter value $\theta_0$, $\operatorname{Pr}_a$ matches the CCP $\operatorname{P}_a$, i.e., $\operatorname{P}_a(v\mid N_t,S_t)=\operatorname{Pr}_a(v\mid (N_{a't})_{a'\in \mathcal{N}_a\cup \{a\}},S_{mt};\theta_0)$.

The vector of parameters $\theta$ contains the parameters entering the mean marginal profits and mean attention that relate to both the covariates and the reference groups $\mathcal{N}_{a}$, $a\in\mathcal{A}$. Note that we assume that agents that are representing the same firm are of the same type. That is, the mean profit and attention functions do not vary across markets for a given firm, but are allowed to be different for different firms. Finally, we do not assume anything about $\mathcal{N}_a$ except that only agents operating within the same market might be in $\mathcal{N}_a$.

\subsection{Estimation} 
\paragraph{Data}
The data we have consist of three objects: (i) the exact date of restaurant openings and closures $\{t_k\}_{k=1}^K$; (ii) the state of the market structure $\{N_{at_k}\}_{a\in\mathcal{A},k=1,\dots,K}$ sampled from a continuous time over interval $[0,t_K]$, where $N_{at_k}$ is the number of restaurants owned by firm $f$ in market $m$ immediately prior to $k$-th change at time $t_k$ ---the last date of measurements coincides with the last day in which any action was observed; and (iii) observable market characteristics $\{S_{m,t_k}\}_{m\in\mathcal{M},k=1,\dots,K}$.

The dataset records the full history of store openings and closures from 1979 through 2024. Each observation corresponds to a discrete event in which a firm opens or closes one or more restaurants in a given city. In total, the data contain 39,198 such events, including 31,509 openings and 7,689 closures, observed over 6,172 distinct action dates. The action dataset was purchased from CnOpenData, a data marketing company.\footnote{\url{https://www.cnopendata.com}}

We use the early portion of the data, from 1979 to 2002, to construct initial conditions for the empirical analysis. These observations allow us to recover the pre-sample distribution of firm presence across markets and to initialize the state variables governing the number of restaurants operated by each firm. The main estimation sample covers the period from 2003 to 2019. Data after 2019 are excluded to avoid confounding effects associated with the COVID-19 pandemic.

To characterize market conditions, we merge the action data with a panel of city-level economic and demographic variables covering the years 2003-2019 obtained from China Data Online.\footnote{\url{https://www.china-data-online.com}} These variables include population, land area, real GDP (deflated to 2019 prices\footnote{Source: World Development Indicators. The last updated date is 2025-12-16.}), and an indicator for provincial capital status. In the overlap period (2003--2019), 63.42\% of days have at least one opening.

\paragraph{Likelihood Function} Our identification argument does not require any parametric assumptions if there is full variation in the size of peer groups. However, for small and moderate-sized samples, we suggest using the parametric maximum likelihood estimator of CCPs $\operatorname{P}_a$, as it is reasonably flexible in allowing market-specific directed network for competition and efficiently uses all variations across markets. We also add the network links in the parametrization of CCPs to estimate the CCPs, these network links, and the other selection and payoff parameters in one step ---instead of first estimating the CCPs and then applying our identification argument to estimate $\mathcal{N}_a$. In particular, we construct from the data a state vector $r_{t_k}=(r_{at_k})_{a\in\mathcal{A}}$, where $r_{at_k}$ indicates whether there is a change in the number of restaurants of firm $f$ in market $m$ at time $t_k$, i.e.,
$
r_{at_k}=2\Char{N_{at_{k+1}}>N_{at_k}}+\Char{N_{at_{k+1}}=N_{at_k}}.
$ The probability of observing $r_{t_k}$, given the data and model parameters $\theta$ conditional on an alarm clock going off, is 
\begin{align*}
p(r_{t_k},S_{t_k},N_{t_k};\theta) &=\prod_{v\in\{0,1,2\}}\prod_{a:r_{at_k}=v} \operatorname{Pr}_a(v\mid (N_{a't})_{a'\in \mathcal{N}_a\cup \{a\}},S_{mt};\theta).
\end{align*}
Hence, the probability that no new restaurants are opened in any market by any firm, given market characteristics and number of restaurants already opened (probability of picking the default), is
\begin{align*}
p_{0}(S_{t_k},N_{t_k};\theta) &=
\prod_{a\in\mathcal{A}}\operatorname{Pr}_a(1\mid (N_{a't})_{a'\in \mathcal{N}_a\cup \{a\}},S_{mt};\theta).
\end{align*}
Finally, given that the arrival process is exponential, the log-likelihood of observing the data given $\theta$ and normalizing $\lambda_a = 1$ (the choice of ``doing nothing'' is not observed) is
\begin{align*}
    \hat{L}(\theta)=\sum_{k=1}^{K} -(t_{k+1}-t_{k})\lambda (1-p_{0}(S_{t_k},N_{t_k};\theta))+\ln (\lambda p(r_{t_{k}},S_{t_k},N_{t_k};\theta)).
\end{align*}
The maximum likelihood estimator of $\theta$, $\hat{\theta}$, is defined as the maximizer of $\hat{L}$ on the parameter space $\Theta$. 

Since the network might be directed (e.g., Wallace ignores KFC but KFC pays attention to Wallace in market $m$), a no restriction network has $(5^2-5)\times281=5,620$ links. That is, there are $2^{5,620}$ possible networks. Checking all possible network structures is not feasible in our application. We use a variation of a greedy algorithm. In step 1 of the algorithm, we start from the network where the link between two firms in a given market is set to zero if and only if at least one of these firms never opened or closed a restaurant in the market. In step 2, we conduct optimization with respect to the rest of the parameters. In step 3, we set one link at a time to zero and find the one link which shutdown leads to the largest improvement in the objective function value. We repeat this step until no link removal leads to an improvement. In step 4, we optimize with respect to the rest of the parameters. Next we alternate between steps 3 and 4 until there are no improvements in the objective function value.

The estimator $\hat{\theta}$ leads to an estimator of CCPs
\[
\hat{\operatorname{P}}_a(v\mid N_t,S_t)=\operatorname{Pr}_a(v\mid (N_{a't})_{a'\in \mathcal{N}_a\cup \{a\}},S_{mt};\hat\theta).
\]
The construction of confidence sets for the parameters and estimated CCPs would require taking into account the estimation error in the estimated network. We leave this difficult problem for future research. An important feature stemming from Assumptions~\ref{ass: MM} and~\ref{ass: simple}, as shown by Proposition~\ref{prop: existence}, is that our model has a single equilibrium or invariant distribution, i.e., the multiplicity of equilibria in the data-generating process is not an issue in our estimation.

\paragraph{Parameterization} We assume that $\operatorname{F}_{\tilde\varepsilon}$ and $\operatorname{F}_{\varepsilon}$ are Logistic c.d.f. Given that our sample size is small relative to the number of agents ---there are $4020$ different dates in which we observed firms opening or closing a restaurant\footnote{For some days, we observe multiple agents opening or closing. Our identification results are still valid if agents have perfectly synchronized clocks.} for $281\times5$ agents--- we use the following second-degree polynomial parameterization to flexibly approximate mean marginal profits and mean attention index: 
\begin{align*}
    \bar{\tilde{\pi}}_{f}(N_{at},N_{a't},S_{mt};\theta)=&S_{mt}\tr \tilde\beta_{f}+\tilde\alpha_{1f}\log(1+N_{(f,m)t})+\tilde\alpha_{2f}\log(1+N_{(f,m)t})^2\\
    &+\tilde\alpha_{3f}\log(1+N_{(f',m)t})+\tilde\alpha_{4f}\log(1+N_{(f',m)t})^2\\
    &+\tilde\alpha_{5f}\left[\log(1+N_{(f,m)t})-\log(1+N_{(f',m)t})\right]^2, \\
    \bar{\pi}_{f}(N_{at},\sum_{a'\in\mathcal{N}}N_{a't},S_{mt};\theta)=&S_{mt}\tr \beta_{f}+\alpha_{1f}\log(1+N_{(f,m)t})+\alpha_{2f}\log(1+N_{(f,m)t})^2\\
    &+\alpha_{3f}\log(1+\sum_{a'\in\mathcal{N}}N_{a't}).  
\end{align*}
The mean marginal profit has two parts: the first one captures the impact of the observable market characteristics, and the second part captures the impact of the own number of restaurants and the number of restaurants of considered competitors.\footnote{We use a second-degree polynomial in our approximation to capture potential nonmonotonicities.} 

\begin{rem}
    Since we use the total number of restaurants opened by each firm in every market as the determinant of selection and expansion probabilities, formally we have a model with infinite history dependence. This, however, does not constitute any issues in our application, since the mean attention and mean marginal profits take known parametric forms. 
\end{rem}

\subsection{Evidence of Limited Attention to Competitors}
\noindent To provide direct evidence that firms do not pay attention to all competitors in every market, we estimate a series of logistic regressions using the daily store-opening data.

\paragraph{Specification} For each focal firm $f$ operating in market $m$, we construct a binary outcome $y_{(f,m)t} \in \{0,1\}$ equal to one if firm $i$ opens at least one store in market $m$ on day $t$, and zero otherwise. The explanatory variables are market characteristics $S_{mt}$ (excluding the indicator for provincial capital status), number of own and competitors restaurants $\log(1+N_{(f,m)t})$ and $\log(1+N_{(f',m)t})$ for firm $f'$ present in market $m$.\footnote{This specification mirrors the preference stage of the structural model: $\log(1+N)$ transformations for store counts and the same five market covariates. The indicator for provincial capital status is excluded because it is constant within a market throughout the sample period and therefore not identified in a market-by-market regression. For three cities (Longnan, Zhongwei, Zunyi) that lack 2003 market covariates, we use the earliest available year, following the same fallback used in the structural estimation.}
The coefficient in front of $\log(1+N_{(f',m)t})$ captures the extent to which firm $f$'s opening decision responds to the cumulative presence of competitor $f'$. 

\paragraph{Sample and estimation}
We estimate the above logistic regression separately for each active firm-market pair, retaining pairs with at least five observed openings during the sample period. Of the remaining pairs, we first attempt estimation with the full set of five covariates; if convergence fails---which occurs in very small markets where GDP, population, and land area are nearly collinear---we re-estimate dropping land area and retaining four covariates (GDP per capita, population density, GDP, and population). The final sample comprises 702 regressions spanning 255 markets and all five chains. Competitors that are never present in a given market are excluded from that
market's regression, following the same convention used in the structural model. The 702 successful regressions are distributed unevenly across chains, reflecting differences in market penetration: Wallace (217 markets), KFC (196), Dicos (136), McDonald's (84), and Burger King (69). The lower coverage for McDonald's and Burger King is consistent with their later and more geographically selective entry relative to the domestic chains and KFC.

\paragraph{Joint likelihood ratio test} First, we test the null hypothesis of ignoring all competitors (i.e., all coefficients in front of $\log(1+N_{(f',m)t})$ are zeros) using a likelihood ratio test. Table~\ref{tab:rf_lrt} reports the percentage of markets in which this
joint null is rejected at the 5\% level.

\begin{table}[ht]
\centering
\caption{Joint likelihood ratio test: percentage of markets rejecting
         $H_0$ (all competitor coefficients zero) at the 5\% level}
\label{tab:rf_lrt}
\begin{tabular}{lcc}
\hline\hline
Firm         & \% markets rejecting & \# markets \\
\hline
Wallace      & 37.0\%               & 216        \\
KFC          & 38.3\%               & 196        \\
McDonald's   & 40.5\%               & 84         \\
Burger King  & 35.3\%               & 68         \\
Dicos        & 38.0\%               & 137        \\
\hline\hline
\end{tabular}
\end{table}

This null is not rejected in 60-65\% of markets, depending on the chain. That is, across the majority of markets, a firm's opening decisions show no statistically significant response to any competitor. However, in 35-40\% of markets, there is still response to actions of competitors. Under our assumption that the firms's parameters do not vary across markets, this indicates that the actions of competitors have effect on marginal profits and the observed variation is coming from limited attention. 

\paragraph{Pairwise rejection rates} Next, we analyze the results for individual coefficient estimates. Table~\ref{tab:rf_pval} reports, for each pair of firms, the percentage of markets in which the null hypothesis that the competitor coefficient is zero is rejected at the 5\% level.

\begin{table}[ht]
\centering
\caption{Percentage of markets with $p$-value $\leq 0.05$ for competitor coefficient
         (row = focal firm, column = competitor)}
\label{tab:rf_pval}
\begin{tabular}{lccccc}
\hline\hline
              & Wallace & KFC    & McDonald's & Burger King & Dicos  \\
\hline
Wallace       & ---     & 16.4\% & 16.7\%     & 15.2\%      & 9.6\%  \\
KFC           & 16.1\%  & ---    & 16.4\%     & 14.5\%      & 15.2\% \\
McDonald's    & 6.0\%   & 25.0\% & ---        & 13.7\%      & 5.3\%  \\
Burger King   & 9.1\%   & 7.4\%  & 12.5\%     & ---         & 13.4\% \\
Dicos         & 14.4\%  & 18.7\% & 9.1\%      & 5.0\%       & ---    \\
\hline\hline
\end{tabular}
\end{table}

Several patterns are consistent with limited attention. McDonald's responds significantly to Dicos in only 5.3\% of markets, and Wallace responds to Dicos in only 9.6\% of markets, suggesting that these chains largely ignore Dicos. Similarly, Dicos responds to Burger King in only 5.0\% of markets, and Burger King responds to KFC in only 7.4\%. By contrast, KFC and McDonald's respond to each other in roughly 16-25\% of markets.

Overall, the results in this section indicate two important observations. First, the actions of competitors affect expansion decisions in a sizable fraction of markets. Second, firms seem to display limited attention to actions of their competitors in many markets. The objective of the structural model that we estimate is to separate these two channels affecting expansion decisions.

\subsection{Estimation Results for the Main Model}

\paragraph{Structure of the Estimated Selection Network}
We begin by summarizing the structure of original and the estimated firm-to-firm selection network.

\begin{figure}[htbp]
\centering
\includegraphics[width=0.45\linewidth]{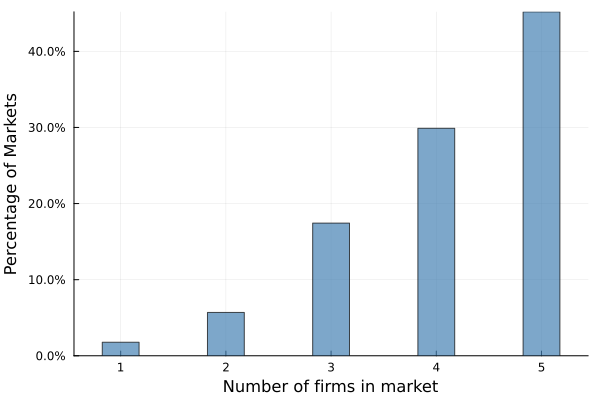}
\caption{Distribution of markets by number of firms present.}
\label{fig:markets-by-firms}
\end{figure}

The initial network that takes into account only firms that were active in the market had $3,874$ links. Figure~\ref{fig:markets-by-firms} depicts the percentage of markets as a function of the number of firms active in them. There is substantial heterogeneity in the number of firms present in the markets. 

\begin{figure}[htbp]
\centering
\includegraphics[width=0.45\linewidth]{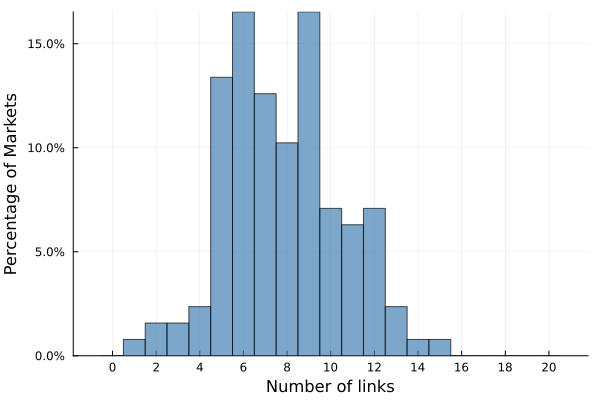}
\caption{Distribution of the number of selection links in markets where all five firms are present. Maximum possible is $20$.}
\label{fig:links-5firm}
\end{figure}

The estimated selection network has $1,547$ directed links. So we use the likelihood value to close down $2,327$ links. Figure~\ref{fig:links-5firm} shows the distribution of the number of such selection links across the markets where all 5 firms are present (there are 127 such markets). Given that the maximal number of links in such a market is 20, the networks for markets with all 5 firms present is rather sparse with the mean across market of $7.82$ (standard deviation is $2.67$). 

\begin{figure}[htbp]
\centering
\includegraphics[width=0.45\linewidth]{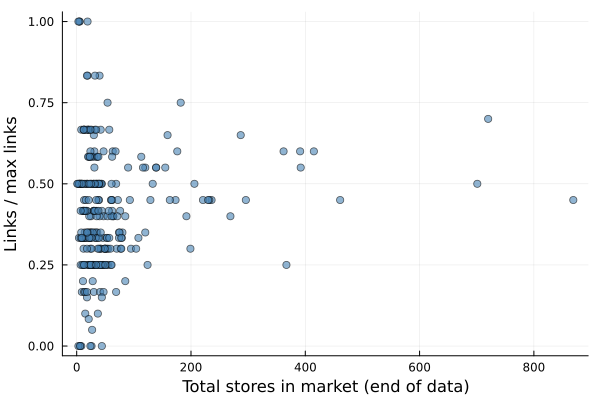}
\caption{Total number of stores in the market (end of data) vs.\ ratio of selection links to maximum possible links in the market. Each point is a market with at least two firms present.}
\label{fig:scatter-stores-link-ratio}
\end{figure}

Next, we study if there is any relation between the size of the market and how attentive firms are. 
Figure~\ref{fig:scatter-stores-link-ratio} plots, for each market with at least two firms, the total number of stores at the end of the data (horizontal axis) against the ratio of the number of selection links in that market to the maximum possible (which equals $k(k-1)$ when $k$ firms are present). Larger markets tend to have more links in levels but the link ratio varies widely.

\begin{figure}[htbp]
\centering
\small
\begin{subfigure}[b]{0.19\linewidth}
  \includegraphics[width=\linewidth]{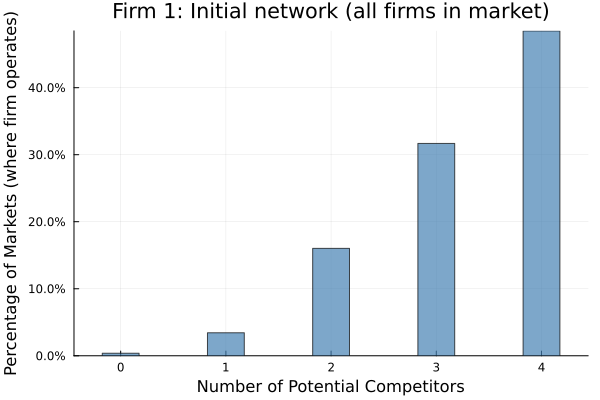}
  \caption{Wallace}
\end{subfigure}\hfill
\begin{subfigure}[b]{0.19\linewidth}
  \includegraphics[width=\linewidth]{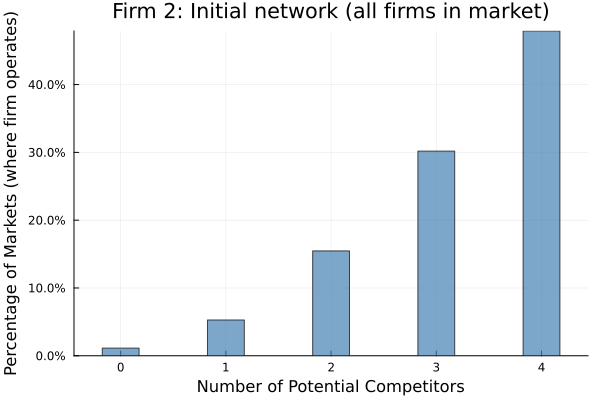}
  \caption{KFC}
\end{subfigure}\hfill
\begin{subfigure}[b]{0.19\linewidth}
  \includegraphics[width=\linewidth]{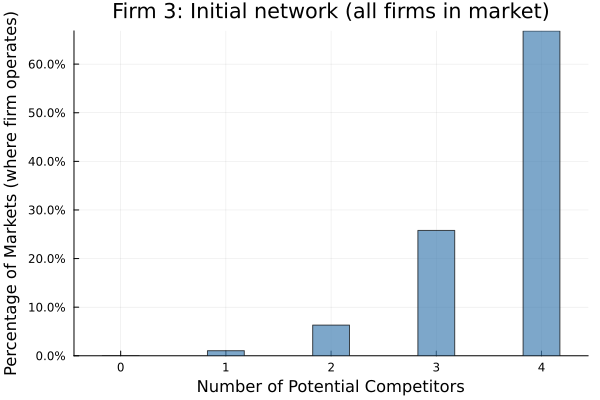}
  \caption{McDonald's}
\end{subfigure}\hfill
\begin{subfigure}[b]{0.19\linewidth}
  \includegraphics[width=\linewidth]{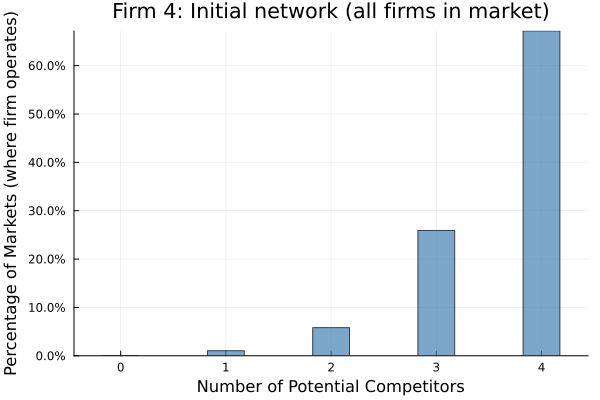}
  \caption{Burger King}
\end{subfigure}\hfill
\begin{subfigure}[b]{0.19\linewidth}
  \includegraphics[width=\linewidth]{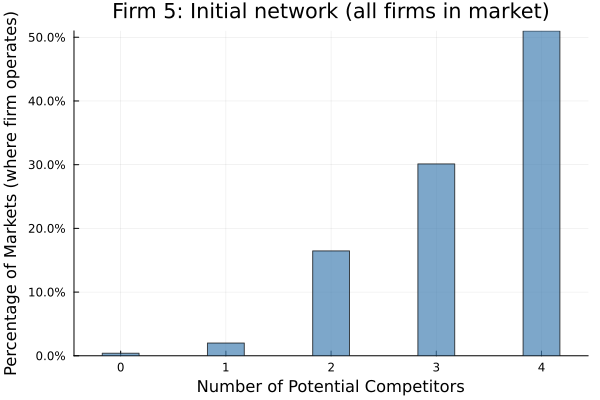}
  \caption{Dicos}
\end{subfigure}\\[0.8em]
\begin{subfigure}[b]{0.19\linewidth}
  \includegraphics[width=\linewidth]{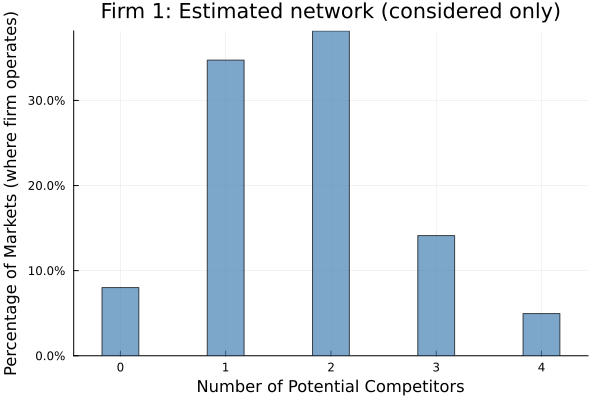}
  \caption{Wallace}
\end{subfigure}\hfill
\begin{subfigure}[b]{0.19\linewidth}
  \includegraphics[width=\linewidth]{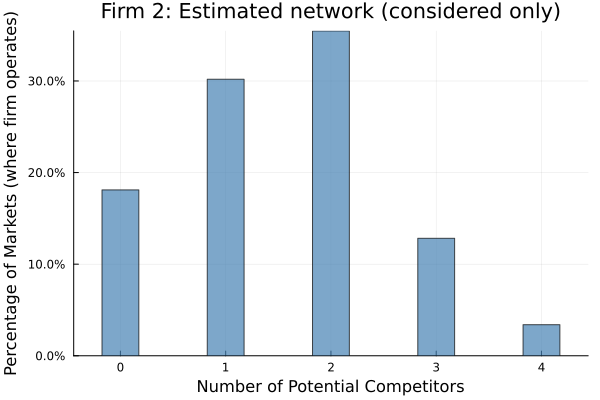}
  \caption{KFC}
\end{subfigure}\hfill
\begin{subfigure}[b]{0.19\linewidth}
  \includegraphics[width=\linewidth]{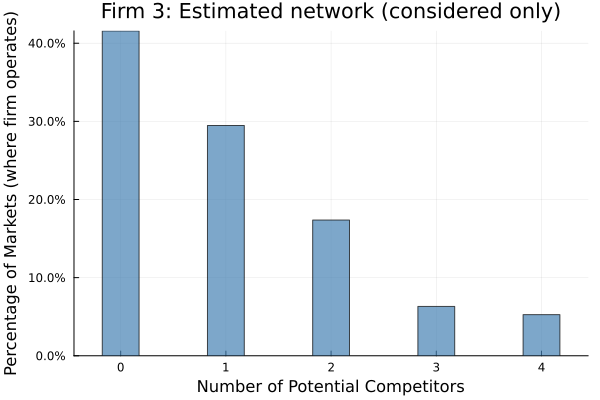}
  \caption{McDonald's}
\end{subfigure}\hfill
\begin{subfigure}[b]{0.19\linewidth}
  \includegraphics[width=\linewidth]{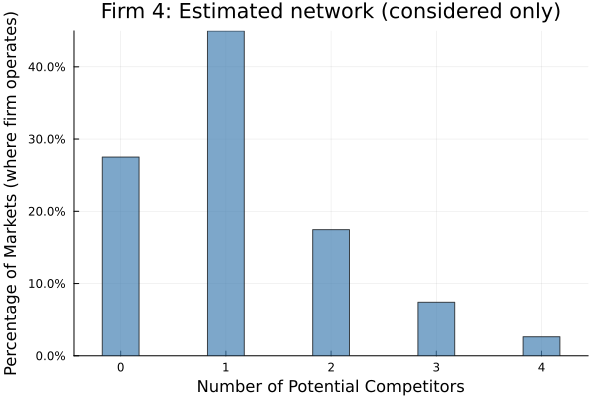}
  \caption{Burger King}
\end{subfigure}\hfill
\begin{subfigure}[b]{0.19\linewidth}
  \includegraphics[width=\linewidth]{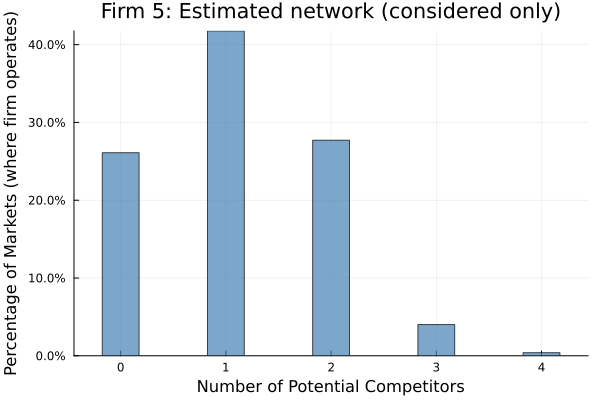}
  \caption{Dicos}
\end{subfigure}
\caption{The distribution of the number of competitors considered with positive probability across markets in the initial-guess network (top row) and the estimated network (bottom row).}
\label{fig:consideration-networks}
\end{figure}

Next, we plot the distribution of the number of  competitors considered with positive probability in markets where the firm operates. In the \emph{initial-guess network} we assumed that every firm considers every other firm that has operated in it; the \emph{estimated network} counts competitors based in estimated $\mathcal{N}_a$. Figure~\ref{fig:consideration-networks} shows both (top row: initial-guess; bottom row: estimated). The shift from initial to estimated reflects limited attention: firms often consider fewer competitors than the present in the market.

\begin{figure}[htbp]
    \centering
    \includegraphics[width=0.8\textwidth]{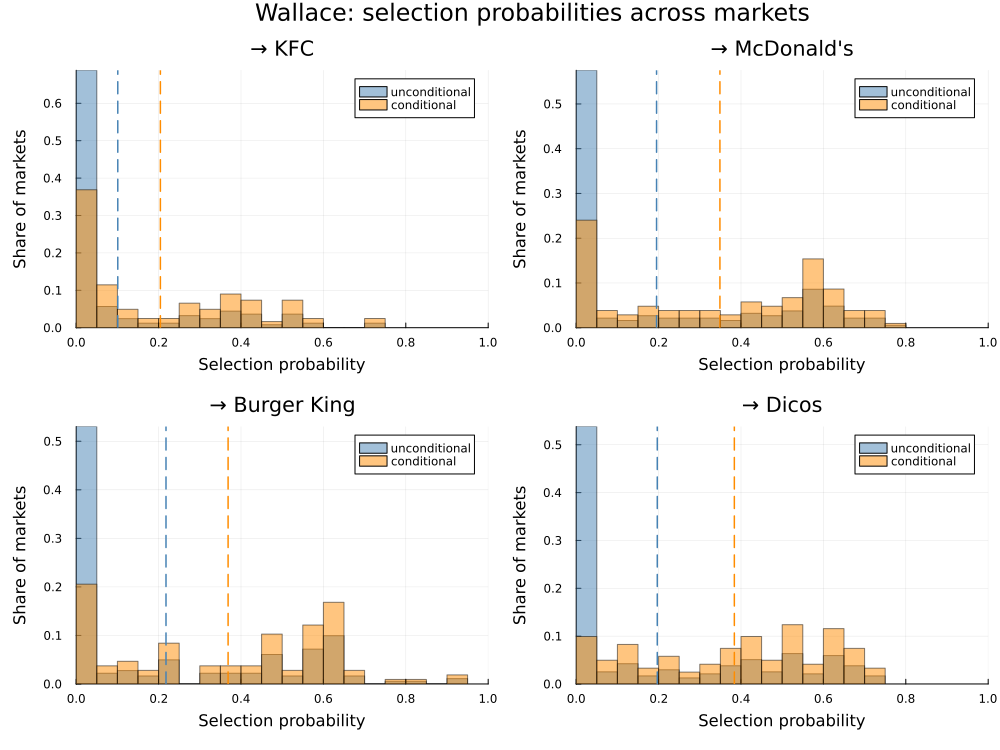}
    \caption{Distribution of selection probabilities for Wallace across markets.
        Each panel shows the histogram for one competitor.
        Blue (unconditional): all markets where both firms are active, with ignored
        pairs counted as zero.
        Orange (conditional): markets where the competitor is not ignored.
        Dashed vertical lines mark the respective means.}
    \label{fig:selprob_wallace}
\end{figure}

\begin{figure}[htbp]
    \centering
    \includegraphics[width=0.8\textwidth]{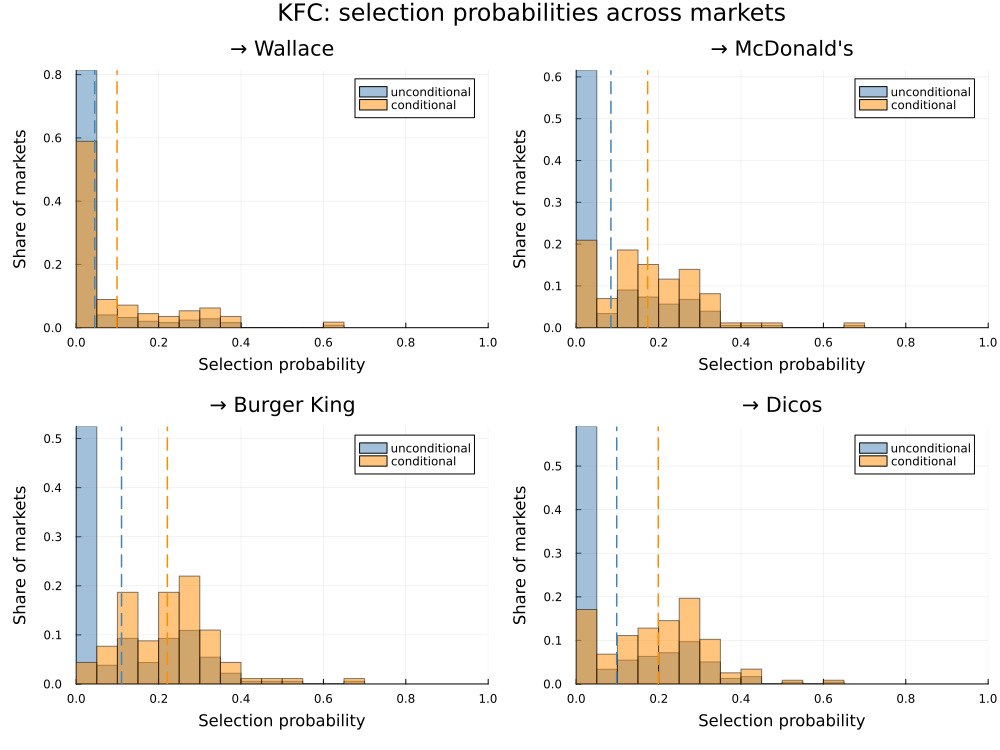}
    \caption{Distribution of selection probabilities for KFC across markets.
        See Figure~\ref{fig:selprob_wallace} for details.}
    \label{fig:selprob_kfc}
\end{figure}

\begin{figure}[htbp]
    \centering
    \includegraphics[width=0.8\textwidth]{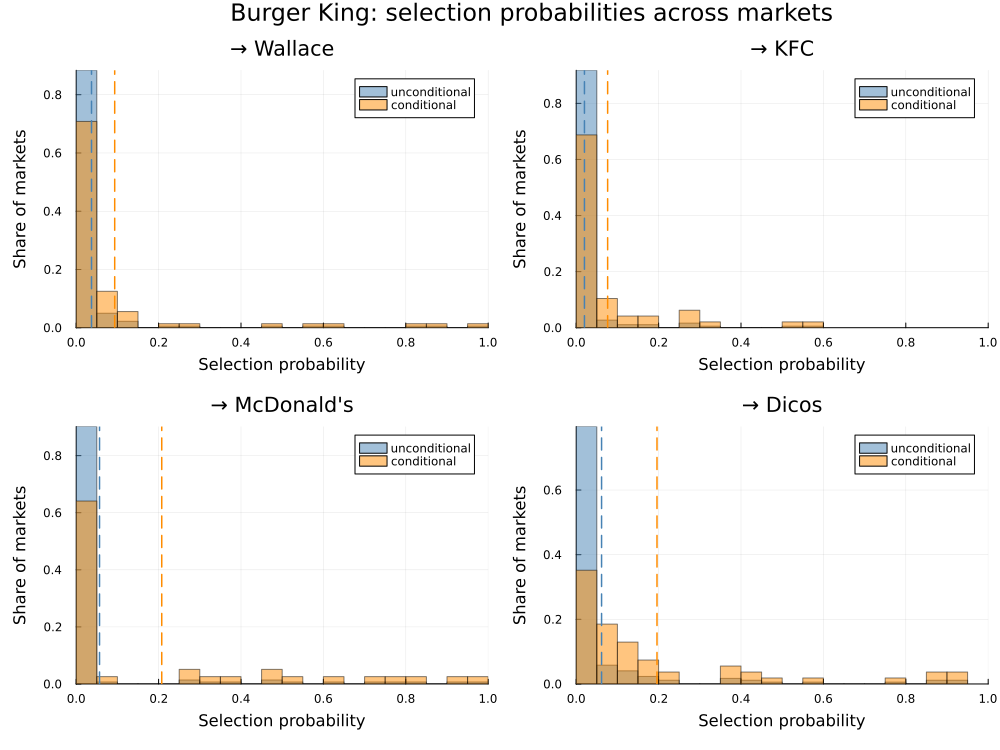}
    \caption{Distribution of selection probabilities for Burger King across markets.
        See Figure~\ref{fig:selprob_wallace} for details.}
    \label{fig:selprob_burgerking}
\end{figure}

\begin{figure}[htbp]
    \centering
    \includegraphics[width=0.8\textwidth]{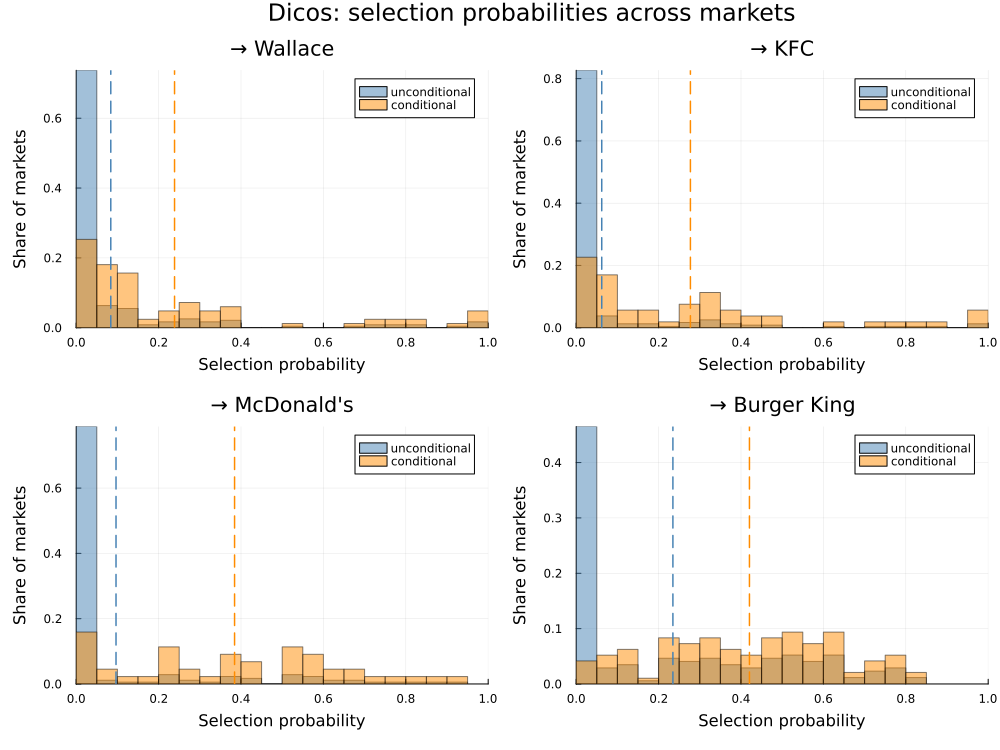}
    \caption{Distribution of selection probabilities for Dicos across markets.
        See Figure~\ref{fig:selprob_wallace} for details.}
    \label{fig:selprob_dicos}
\end{figure}

\paragraph{Selection of Competitors} We calculated the selection probabilities for every market and each directed firm pair (e.g, the probability of Wallace selecting KFC in Beijing) using the numbers of restaurants and covariate values observed at the end of the measurements used in estimation (December 31, 2019). Figures~\ref{fig:selprob_wallace}~-~\ref{fig:selprob_dicos} depict the distribution of these probabilities across all markets where these firms are present (blue) and conditional on the target firm being considered (orange) for all firms except McDonald's since its selection probabilities are almost everywhere equal to zero. These figures show that firms ignore many of their competitors in many markets. However, conditional on selecting a firm with positive probability (i.e., $a'\in\mathcal{N}_a$), all firms except McDonalds display huge heterogeneity in paying attention to competitors. Overall, firms display striking differences: McDonald's is essentially not considering anyone as competitors, while other firms display reasonable attention with their average selection probabilities conditional on considering a firm varying in the range of $0.1$-$0.4$ in 2019.\footnote{There are could be other factors affecting firms' expansion decisions that are not captured by the model. For example, a firm may operate under a different corporate strategy.}

\begin{figure}[htbp]
\centering
\small
\begin{subfigure}[b]{0.3\linewidth}
  \includegraphics[width=\linewidth]{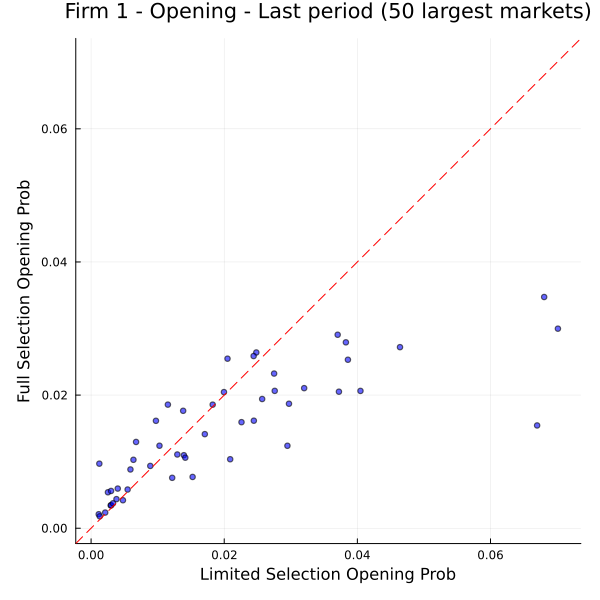}
  \caption{Wallace}
\end{subfigure}\hfill
\begin{subfigure}[b]{0.3\linewidth}
  \includegraphics[width=\linewidth]{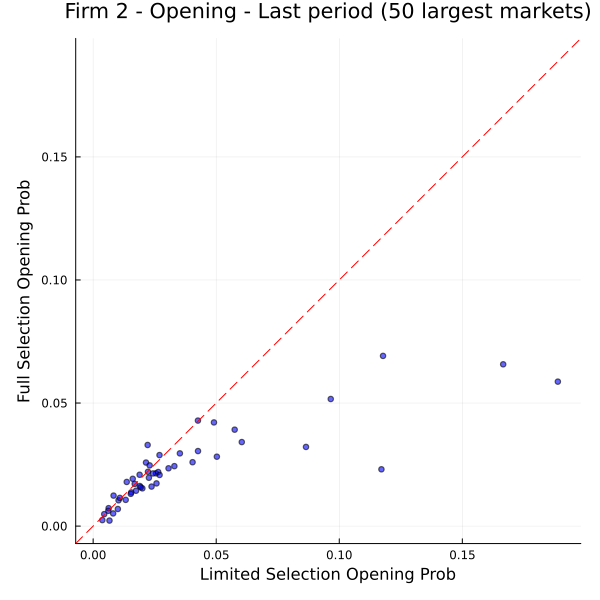}
  \caption{KFC}
\end{subfigure}\hfill
\begin{subfigure}[b]{0.3\linewidth}
  \includegraphics[width=\linewidth]{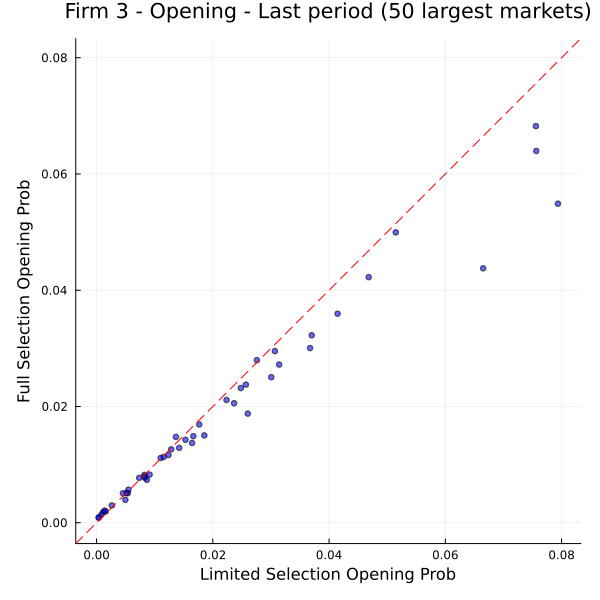}
  \caption{McDonald's}
\end{subfigure}\hfill
\begin{subfigure}[b]{0.3\linewidth}
  \includegraphics[width=\linewidth]{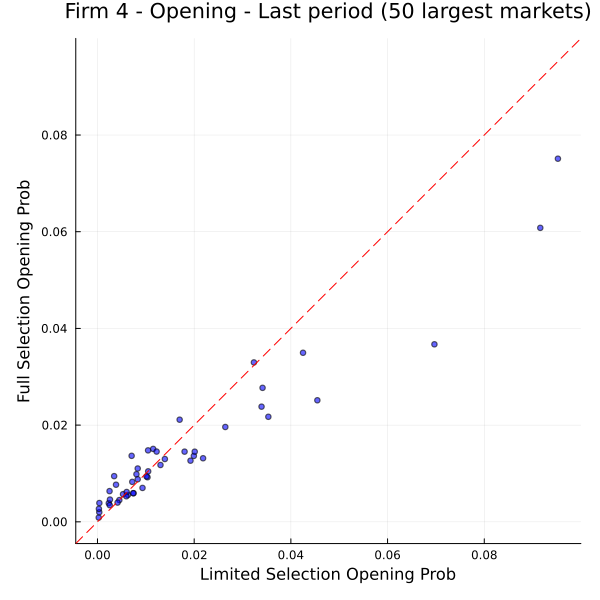}
  \caption{Burger King}
\end{subfigure}\hfill
\begin{subfigure}[b]{0.3\linewidth}
  \includegraphics[width=\linewidth]{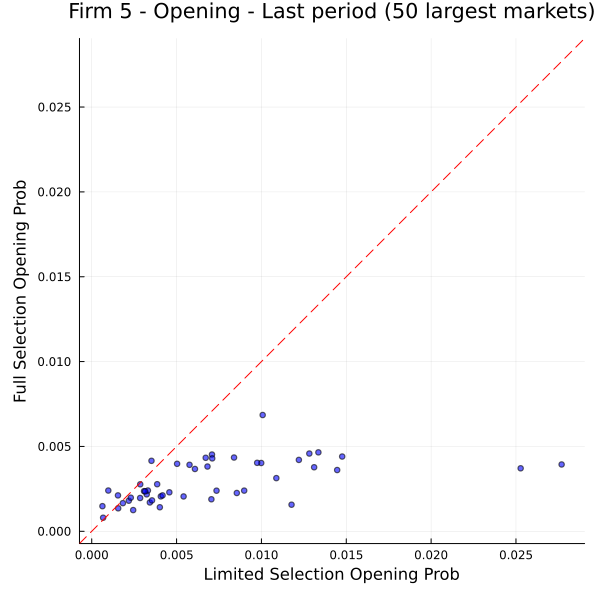}
  \caption{Dicos}
\end{subfigure}
\caption{Opening probability: limited vs.\ full selection, 50 largest markets, last period.}
\label{fig:full-vs-limited-opening-top50}
\end{figure}

\paragraph{Marginal Profits Estimates} We analyze the probabilities of opening a new restaurant across $50$ largest in terms of the number of restaurants markets. To quantify the effect of adding limited selection of competitors, we also estimated the marginal profit parameters assuming that all competitors that ever operated in the market are considered. We refer to the former as limited selection estimates and to the latter as full selection estimates. The results of the estimation of opening probabilities  are presented in Figure~\ref{fig:full-vs-limited-opening-top50}. Table~\ref{tab: full vs.limited. stats} reports different statistics for these scatter plots: the difference in average across markets probabilities, the slope of regressing full selection probabilities on limited selection probabilities and a constant, sample correlation, and the share of markets above 45-degree line. The average difference captures the average level shift induced by forcing firms to pay attention to all competitors. If it has a positive sign, then full selection increases opening probabilities on average, while a negative sign indicates the opposite. The slope measures how strongly full selection responds to cross market variation in limited selection probabilities. A slope close to one implies that market rankings (in terms of opening probabilities) are invariant and differences are mainly level shifts. A slope well below one indicates attenuation: full selection compresses dispersion, pulling high probability markets down and reducing cross market differences. The correlation coefficient captures whether the ranking of markets changes. A correlation close to one means that limited and full selection rank markets similarly, while lower correlation indicates reordering. Finally, the share of markets above the 45 degree line shows in how many markets full selection predicts more expansion than limited selection, providing a simple directional measure of competitive pressure.

The values of these statistics display several interesting patterns across firms. First, limited attention systematically inflates expansion probabilities since for all firms, the mean difference between full and limited attention is negative. McDonald's continues to display the strongest alignment between the two specifications, with a high correlation of 0.98 and a slope close to 0.80, implying that full attention largely preserves the ranking of markets while reducing expansion levels proportionally. Burger King exhibits small attenuation with small reordering. Wallace and KFC display stronger compression effects, with slopes of 0.40 and 0.32, respectively. This means that full attention substantially reduces cross market dispersion in opening probabilities. However, correlations above 0.80 suggest that market rankings remain mostly the same. In stark contrast, Dicos has a slope of only 0.12 and a lower correlation of 0.55. This indicates both severe compression and reordering of market attractiveness under full attention. Overall, the estimates show that limited attention generates systematic overexpansion for all firms, but the magnitude and nature of the distortions vary: for most firms it primarily shifts levels and attenuates dispersion, while for Dicos it fundamentally reshapes cross market responsiveness.

\begin{table}[htbp]
\centering
\caption{Opening Probabilities: Full vs Limited selection (Top 50 Markets)}
\label{tab: full vs.limited. stats}
\small
\begin{tabular}{lcccc}
\toprule
Firm & $Average_{Full}-Average_{Limited}$ & Slope & Correlation & Share Above 45$^\circ$ \\
\midrule
Wallace      & -0.00514 & 0.3981 & 0.8034 & 0.5 \\
KFC          & -0.01283 & 0.3223 & 0.8441 & 0.3061 \\
McDonald's   & -0.00266 & 0.7946 & 0.9842 & 0.2979 \\
Burger King  & -0.00284 & 0.6465 & 0.9645 & 0.4583 \\
Dicos        & -0.00402 & 0.1208 & 0.5500 & 0.102 \\
\bottomrule
\end{tabular}
\end{table}


Taken together, the estimates suggest that heterogeneity in selection networks translates directly into heterogeneity in how limited selection distorts expansion behavior. For firms that do not pay much attention to their competitors, limited selection primarily shifts levels. For firms with large attention networks, limited selection alters both levels and cross market responsiveness. This findings emphasize the importance of limited selection and peer group heterogeneity in shaping expansion dynamics.

\section{Concluding Remarks}\label{sec: conclusion}

\noindent This paper studies dynamic interactions among agents that are connected through a social network. In the model, each agent is linked to a finite set of agents and selects an option from a finite set of alternatives. The timing of the decision making of each group member follows a simple Poisson process that is independent across the agents. The distinctive feature of the model is that, at the moment of making a decision, the agent does not pay attention to all her linked agents. Instead, she first selects a subset of peers and then makes a decision under their influence. The previous choices of the linked agents affect both the probability that different agents are included in her reference group and the preferences of the agent over the alternatives once the group is formed. This model can lead to choice-based homophily in decision-making. 

We exploit variation in the choices of all agents through time and variation in the size of their reference groups to recover all parts of the model. These parts include random preferences and the probability of paying attention to different agents. 

The application of our framework to the analysis of the opening and closing decisions of the five largest fast-food restaurants in China shows that heterogeneity in peer selection networks translates into systematic differences in expansion dynamics between firms. Limited selection generates excess adjustment and, for some firms, substantial distortion in cross market responsiveness, emphasizing the economic importance of modeling endogenous and directed attention in competitive environments.

\bibliography{references}
\appendix
\section{Proofs}\label{app: proofs}
\subsection{Proof of Proposition~\ref{prop: existence}}
\noindent \noindent For an irreducible, finite-state, continuous Markov chain, the equilibrium $\mu$ exists, is unique and has full support. Note that
\begin{equation*}
\operatorname{P}_{a}\left( v \mid \mathbf{y}\right) =\sum\nolimits_{\mathcal{N} \subseteq \mathcal{N}_{a}}\operatorname{R}^{a}\left(v \mid \mathbf{y}, \mathcal{N}\right)\operatorname{S}^{a}\left(\mathcal{N} \mid \mathbf{y}, \mathcal{N}_a \right).
\end{equation*}
\noindent Note that $\operatorname{P}_{a}\left( v \mid \mathbf{y}\right)$ is a weighted average of $1>\operatorname{R}^{a}\left(v \mid \mathbf{y}, \mathcal{N}\right)>0$. Thus, $1>\operatorname{P}_{a}\left( v \mid \mathbf{y}\right)>0$
for all $a$ and $\mathbf{y}$, and we can go from one configuration to the other one in less than $A$ steps with a positive probability.

\subsection{Proof of Proposition~\ref{prop: network}}
\noindent Fix any two different Agents $a,a' \in\mathcal{A}$. Let $\mathbf{y}_{a'}^v$ be the choice configuration obtained from $\mathbf{y}$ by replacing the $a'$ component, $y_{a'}$, by $v$. First, note that if $a'\not\in\mathcal{N}_a$ then 
\[
\operatorname{P}_a(v\mid\mathbf{0}^v_{a'})-\operatorname{P}_a(v\mid\mathbf{0})=0,
\]
where $\mathbf{0}=(0,\dots,0)\tr$ is the choice configuration where everyone picks $0$. We next show that if the previous difference in CCPs is different from zero, then $a'\in\mathcal{N}_a$. Note that if $a'\in\mathcal{N}_a$
\begin{align*}
    &\operatorname{P}_a(v\mid\mathbf{y})=\operatorname{Q}^{a}(a'\mid \mathbf{y})\sum_{\mathcal{N}\subseteq\mathcal{N}_a\setminus\{a'\}}\operatorname{R}^{a}\left(v \mid \mathbf{y}, \mathcal{N}\cup\{a'\}\right){\operatorname{S}}^{a}\left(\mathcal{N} \mid \mathbf{y}, \mathcal{N}_a\setminus\{a'\} \right) +\\
    &(1-\operatorname{Q}^a(a'\mid \mathbf{y}))\sum_{\mathcal{N}\subseteq\mathcal{N}_a\setminus\{a'\}}\operatorname{R}^{a}\left(v \mid \mathbf{y}, \mathcal{N}\right){\operatorname{S}}^{a}\left(\mathcal{N} \mid \mathbf{y}, \mathcal{N}_a\setminus\{a'\} \right)=\\
    &\operatorname{Q}^a(a'\mid \mathbf{y})\sum_{\mathcal{N}\subseteq\mathcal{N}_a\setminus\{a'\}}\left[\operatorname{R}^{a}\left(v \mid \mathbf{y}, \mathcal{N}\cup\{a'\}\right)-\operatorname{R}^{a}\left(v \mid \mathbf{y}, \mathcal{N}\right)\right]{\operatorname{S}}^{a}\left(\mathcal{N} \mid \mathbf{y}, \mathcal{N}_a\setminus\{a'\} \right)+\\
    &\sum_{\mathcal{N}\subseteq\mathcal{N}_a\setminus\{a'\}}\operatorname{R}^{a}\left(v \mid \mathbf{y}, \mathcal{N}\right)\operatorname{S}^{a}\left(\mathcal{N} \mid \mathbf{y}, \mathcal{N}_a\setminus\{a'\} \right).
\end{align*}
Moreover, by Assumption~\ref{ass: simple}(ii), $\operatorname{R}^{a}\left(v \mid \mathbf{y}, \mathcal{N}\right)$ does not depend on $y_{a'}$ for any $\mathcal{N}$ that does not contain $a'$. Similarly, by Assumption~\ref{ass: simple}(i), $\operatorname{S}^{a}\left(\mathcal{N} \mid \mathbf{y}, \mathcal{N}_a\setminus\{a'\} \right)$ does not depend on $y_{a'}$.
Hence, 
\begin{align*}
    &\operatorname{P}_a(v\mid\mathbf{0}^v_{a'})-\operatorname{P}_a(v\mid\mathbf{0})=\\
    &\operatorname{Q}^a(a'\mid \mathbf{0}^v_{a'})\sum_{\mathcal{N}\subseteq\mathcal{N}_a\setminus\{a'\}}\left[\operatorname{R}^{a}\left(v \mid \mathbf{0}^v_{a'}, \mathcal{N}\cup\{a'\}\right)-\operatorname{R}^{a}\left(v \mid \mathbf{0}^v_{a'}, \mathcal{N}\right)\right]{\operatorname{S}}^{a}\left(\mathcal{N} \mid \mathbf{0}^v_{a'}, \mathcal{N}_a\setminus\{a'\} \right)\\
    -&\operatorname{Q}^a(a'\mid \mathbf{0})\sum_{\mathcal{N}\subseteq\mathcal{N}_a\setminus\{a'\}}\left[\operatorname{R}^{a}\left(v \mid \mathbf{0}, \mathcal{N}\cup\{a'\}\right)-\operatorname{R}^{a}\left(v \mid \mathbf{0}, \mathcal{N}\right)\right]{\operatorname{S}}^{a}\left(\mathcal{N} \mid \mathbf{0}, \mathcal{N}_a\setminus\{a'\} \right)=\\
    &\operatorname{Q}^a(a'\mid \mathbf{0}^v_{a'})\sum_{\mathcal{N}\subseteq\mathcal{N}_a\setminus\{a'\}}\left[\operatorname{R}^{a}\left(v \mid \mathbf{0}^v_{a'}, \mathcal{N}\cup\{a'\}\right)-\operatorname{R}^{a}\left(v \mid \mathbf{0}, \mathcal{N}\right)\right]{\operatorname{S}}^{a}\left(\mathcal{N} \mid \mathbf{0}, \mathcal{N}_a\setminus\{a'\} \right)\\
    -&\operatorname{Q}^a(a'\mid \mathbf{0})\sum_{\mathcal{N}\subseteq\mathcal{N}_a\setminus\{a'\}}\left[\operatorname{R}^{a}\left(v \mid \mathbf{0}, \mathcal{N}\cup\{a'\}\right)-\operatorname{R}^{a}\left(v \mid \mathbf{0}, \mathcal{N}\right)\right]{\operatorname{S}}^{a}\left(\mathcal{N} \mid \mathbf{0}, \mathcal{N}_a\setminus\{a'\} \right)\neq 0,
\end{align*}
where the last inequality follows from Assumption~\ref{ass: regularity}.

\subsection{Proof of Proposition~\ref{prop: selection}}
\noindent Fix some type $t\in\mathcal{H}$. Since $\operatorname {N}_t \geq 3$, there are at least 3 agents of type $t$ with different number of peers. Let these agents be $a_1$, $a_2$, and $a_3$ and let $\operatorname {N}_1<\operatorname {N}_2<\operatorname {N}_3$ be the number of agents in their corresponding peer reference groups. Take any three configurations $\mathbf{y}^{j}$, $j=1,2,3$, such that $y^{1}_{a_1}=y^{2}_{a_2}=y^{3}_{a_3}=v^*$ for some $v^*\in\mathcal{Y}$ and all other components are set to $v'$, which may coincide with or be different from $v^*$. Note that 
\[
\operatorname{Q}^{a_1}(a'\mid\mathbf{y}^{1},\mathcal{N}_{a_1})=\operatorname{Q}^{a_2}(a''\mid\mathbf{y}^{2},\mathcal{N}_{a_2})=\operatorname{Q}^{a_3}(a'''\mid\mathbf{y}^{3},\mathcal{N}_{a_3})=\operatorname{Q}_{t}(v^*,v'),
\]
for any $a'\in\mathcal{N}_{a_1}$, $a''\in\mathcal{N}_{a_2}$, and $a'''\in\mathcal{N}_{a_3}$.

Let $q=\operatorname{Q}_{t}(v^*,v')$ and $t(N)=1-(1-q)^{N}$ be the probability that at least one peer out of $N$ peers is considered. Under Assumption~\ref{ass: average}(i), since all 3 agents pick the same option and all their peers are picking the same alternative we have that 
\begin{align*}
\operatorname{P}_{a_j}(v\mid \mathbf{y}^{j})&=\operatorname{R}_t(v\mid v^*, \mathbf{0})(1-t(N_{j}))+\operatorname{R}_{t}(v\mid v^*,\mathbf{0}_{v'}^1)t(N_j)\\
&=\operatorname{R}_{t}(v\mid v^*,\mathbf{0})+[\operatorname{R}_{t}(v\mid v^*,\mathbf{0}_{v'}^1)-\operatorname{R}_{t}(v\mid v^*,\mathbf{0})]t(N).
\end{align*}
 
\noindent Thus, given that $\operatorname{R}_{t}(v\mid v^*,\mathbf{0}_{v'}^1)-\operatorname{R}_{t}(v\mid v^*,\mathbf{0})\neq 0$, we deduce that 
\[
\dfrac{\operatorname{P}_{a_3}(v\mid \mathbf{y}^{3})-\operatorname{P}_{a_1}(v\mid \mathbf{y}^{1})}{\operatorname{P}_{a_2}(v\mid \mathbf{y}^{2})-\operatorname{P}_{a_1}(v\mid \mathbf{y}^{1})}=\dfrac{t(N_3)-t(N_1)}{t(N_2)-t(N_1)}.
\]
Since the left-hand-side of the last expression is observed, if we show that the right-hand-side is a known strictly monotone function of $q$, then we prove that $q$ can be identified from the data. Let $x=1-q$, $n_2 = N_2 - N_1$, and $n_3 = N_3 - N_1$ and note that
\[
f(x) = \dfrac{t(N_3)-t(N_1)}{t(N_2)-t(N_1)} = \dfrac{1-x^{n_3}}{1-x^{n_2}}.
\]
We next show that $f'(x) > 0$ for all $x\in(0,1)$ and $n_3>n_2$. After some manipulation, we get
\[
f'(x) = \dfrac{n_2x^{n_2 - 1} - n_3x^{n_3 - 1}+(n_3 -n_2)x^{n_3+n_2 - 1}}{(1-x^{n_2})^2}.
\]
Note that the denominator is strictly positive and the numerator can be written as
\[
x^{n_2 - 1}(n_2 - n_3x^{n_3 - n_2}+(n_3 -n_2)x^{n_3}).
\]
Note that 
\[
\tilde{f}(x)=n_2 - n_3x^{n_3 - n_2}+(n_3 -n_2)x^{n_3}
\]
is such 
\[
\tilde{f}'(x)=- n_3(n_3 - n_2)x^{n_3 - n_2-1}+(n_3 -n_2)n_3x^{n_3-1}=(n_3 -n_2)n_3x^{n_3 - n_2-1}[x^{n_2}-1]<0
\]
for all $x\in(0,1)$. Hence, $\tilde{f}$ is strictly decreasing on $(0,1)$ and thus $\tilde{f}(x)\geq \tilde{f}(1)=0$  for all $x$. Thus, $f'(x)>0$ for all $x\in(0,1)$ and $f$ is strictly increasing.
%
Since, $x=1-q$, then $q$ is identified from observed CCPs. 

Next, we identify $t(N)$ from $q$ and 
\[
\operatorname{R}_{t}(v\mid v^*,\mathbf{0}_{v'}^1)-\operatorname{R}_{t}(v\mid v^*,\mathbf{0})=\dfrac{\operatorname{P}_{a_3}(v\mid \mathbf{y}^{3})-\operatorname{P}_{a_1}(v\mid \mathbf{y}^{1})}{t(N_3)-t(N_1)}.
\]
Finally, we identify
\[
\operatorname{R}^a(v\mid \mathbf{y},\emptyset)=\operatorname{R}_{t}(v\mid v^*,\mathbf{0})=\operatorname{P}_{a_3}(v\mid \mathbf{y}^{3})-\dfrac{\operatorname{P}_{a_3}(v\mid \mathbf{y}^{3})-\operatorname{P}_{a_1}(v\mid \mathbf{y}^{1})}{t(N_3)-t(N_1)}t(N_3)
\]
and 
\[
\operatorname{R}^a(v\mid \mathbf{y},\{a'\})=\operatorname{R}_{t}(v\mid v^*,\mathbf{0}_{v'}^1)=\operatorname{P}_{a_3}(v\mid \mathbf{y}^{3})+\dfrac{\operatorname{P}_{a_3}(v\mid \mathbf{y}^{3})-\operatorname{P}_{a_1}(v\mid \mathbf{y}^{1})}{t(N_3)-t(N_1)}(1-t(N_3))
\] 
for any $a'\in\mathcal{N}_a$ and $\mathbf{y}$ such that $y_a=v^*$ and $y_{a'}=v'$. The fact that the choice of $v^*$, $v'$, and $t$ was arbitrary completes the proof.

\subsection{Proof of Proposition~\ref{prop: logit rule}}
\noindent Note that 
\[
    \operatorname{R}^a(v\mid\mathbf{y},\emptyset)=\dfrac{e^{\alpha_{h(a),v}(y_a)}}{1+\sum_{v'\in\mathcal{Y}\setminus\{0\}}e^{\alpha_{h(a),v'}(y_a)}}.
    \]
    Hence, $\alpha_{h(a),v}(y_a)=\log(\operatorname{R}^a(v\mid\mathbf{y},\emptyset))-\log(\operatorname{R}^a(0\mid\mathbf{y},\emptyset))$. To identify $\beta_{h(a),v}(y_a)$, note that
    \[
    \operatorname{R}^a(v\mid\mathbf{y},\{a'\})=\dfrac{e^{\alpha_{h(a),v}(y_a)+\beta_{h(a),v}(y_a)\Char{y_{a'}=v}}}{\sum_{v'\in\mathcal{Y}}e^{\alpha_{h(a),v'}(y_a)+\beta_{h(a),v'}(y_a)\Char{y_{a'}=v'}}}.
    \]
    As a result, for $v\neq 0$, 
    \[
    \beta_{h(a),v}(y_a)=\log(\operatorname{R}^a(v\mid\mathbf{y}^*,\{a'\}))-\log(\operatorname{R}^a(0\mid\mathbf{y}^*,\{a'\}))-\alpha_{h(a),v}(y_a),
    \]
    for any $\mathbf{y}^*$ such that $y^*_{a'}=v$. To identify $\beta_{h(a),0}(y_a)$, we can just flip the alternatives:
    \[
    \beta_{h(a),0}(y_a)=\log(\operatorname{R}^a(0\mid\mathbf{y}^*,\{a'\}))-\log(\operatorname{R}^a(v''\mid\mathbf{y}^*,\{a'\}))+\alpha_{h(a),v''}(y_a),
    \]
    for any $\mathbf{y}^*$ such that $y^*_{a'}=0$.

\subsection{Proof of Proposition~\ref{prop: idoutside}}
\noindent Consider Agent $a$ and let $\mathbf{0}^{v_1}_a$ be a vector where all agents select the outside option, except Agent $a$ that is assigned option $v_1$. If Assumption~\ref{ass: outside} holds, then
\begin{equation*}
\operatorname{P}_{a}\left( v \mid \mathbf{0}^{v_1}_a\right) =\operatorname{R}_{a}\left(v \mid \ v_1,\mathbf{0}\right). 
\end{equation*}
That is, we recover the choice rule for the empty set of active peers. Recall that (by Proposition~\ref{prop: network}) we can recover $\mathcal{N}_a$. Consider $a' \in \mathcal{N}_a$ and let $\left(\mathbf{0}^{v_1}_a\right)^{v_2}_{a'}$ be a vector where all agents select the outside option, except Agents $a$ and $a'$ that are assigned options $v_1$ and $v_2$, respectively. Note that for Agent $a'$ that is not always selected 
\[
\operatorname{P}_{a}\left( v \mid \left(\mathbf{0}^{v_1}_a\right)^{v_2}_{a'}\right) =\operatorname{R}_{a}\left(v \mid \ v_1,\mathbf{0}^1_{v_2}\right)\operatorname{Q}_a\left(a'\mid v_1,v_2\right) + \operatorname{R}_{a}\left(v \mid \ v_1,\mathbf{0}\right)\left(1-\operatorname{Q}_a\left(a'\mid v_1,v_2\right)\right). 
\]
Hence,
\begin{equation*}
\operatorname{P}_{a}\left( v \mid \left(\mathbf{0}^{v_1}_a\right)^{v_2}_{a'}\right)-\operatorname{P}_{a}\left( v \mid \mathbf{0}^{v_1}_a\right) =\left[\operatorname{R}_{a}\left(v \mid \ v_1,\mathbf{0}^1_{v_2}\right)-\operatorname{R}_{a}\left(v \mid \ v_1,\mathbf{0}\right)\right]\operatorname{Q}_a\left(a'\mid v_1,v_2\right).
\end{equation*}
Thus, if for $a$, $v_1$, $v_2$, the maximum selection probability is known, then
\[
\max_{a'\in\mathcal{N}_a} \abs{\operatorname{P}_{a}\left( v \mid \left(\mathbf{0}^{v_1}_a\right)^{v_2}_{a'}\right)-\operatorname{P}_{a}\left( v \mid \mathbf{0}^{v_1}_a\right)}=\abs{\operatorname{R}_{a}\left(v \mid \ v_1,\mathbf{0}^1_{v_2}\right)-\operatorname{R}_{a}\left(v \mid \ v_1,\mathbf{0}\right)}\max_{a'\in\mathcal{N}_a} \operatorname{Q}_a\left(a'\mid v_1,v_2\right).
\]
If the minimum selection probability is known, then 
\[
\min_{a'\in\mathcal{N}_a} \abs{\operatorname{P}_{a}\left( v \mid \left(\mathbf{0}^{v_1}_a\right)^{v_2}_{a'}\right)-\operatorname{P}_{a}\left( v \mid \mathbf{0}^{v_1}_a\right)}=\abs{\operatorname{R}_{a}\left(v \mid \ v_1,\mathbf{0}^1_{v_2}\right)-\operatorname{R}_{a}\left(v \mid \ v_1,\mathbf{0}\right)}\min_{a'\in\mathcal{N}_a} \operatorname{Q}_a\left(a'\mid v_1,v_2\right).
\]
Since the selection probability is positive, we identify 
\[
\operatorname{R}_{a}\left(v \mid \ v_1,\mathbf{0}^1_{v_2}\right)-\operatorname{R}_{a}\left(v \mid \ v_1,\mathbf{0}\right).
\]
Hence, we also identify
\[
\operatorname{Q}_a\left(a'\mid v_1,v_2\right)=\dfrac{\operatorname{P}_{a}\left( v \mid \left(\mathbf{0}^{v_1}_a\right)^{v_2}_{a'}\right)-\operatorname{P}_{a}\left( v \mid \mathbf{0}^{v_1}_a\right)}{ \operatorname{R}_{a}\left(v \mid \ v_1,\mathbf{0}^1_{v_2}\right)-\operatorname{R}_{a}\left(v \mid \ v_1,\mathbf{0}\right)}.
\]
Since the choice of $v_1$, $v_2$ was arbitrary, we identify the selection mechanism.

To identify the choice rule, note that since
\[
\operatorname{R}_{a}\left(v \mid \ v_1,\mathbf{0}^1_{v_2}\right)-\operatorname{R}_{a}\left(v \mid \ v_1,\mathbf{0}\right).
\]
and 
\begin{equation*}
\operatorname{R}_{a}\left(v \mid \ v_1,\mathbf{0}\right)=\operatorname{P}_{a}\left( v \mid \mathbf{0}^{v_1}_a\right) 
\end{equation*}
are identified, we also identify $\operatorname{R}_{a}\left(v \mid \ v_1,\mathbf{0}_{v_2}^{1}\right)$. Hence, we identify the choice rule for any singleton $\mathcal{N}$. Then, by switching one by one all other peers to any non-default alternative recursively (similar to the proof of Proposition~\ref{prop: recursive}) we can identify the choice rule for $\mathcal{N}$ of any size.

\end{document}